\documentclass[11pt, oneside]{article}
\usepackage{geometry}
\usepackage{setspace} 
\usepackage[parfill]{parskip}
\usepackage{lineno} 
\usepackage{amssymb}
\usepackage{amsmath}
\usepackage{dsfont}
\usepackage{mathptmx} 
\usepackage{graphicx}
\usepackage[labelfont=bf]{caption} 
\usepackage{float} 
\usepackage[
	style=authoryear,
	citestyle=authoryear,
	maxbibnames=6,
	giveninits=true,
	backend=bibtex,
	url=false,
	doi=false,
	isbn=false,
	eprint=false
	]{biblatex}
\newcommand{\beginsupplement}{%
        \setcounter{table}{0}
        \renewcommand{\thetable}{S\arabic{table}}%
        \setcounter{figure}{0}
        \renewcommand{\thefigure}{S\arabic{figure}}%
     }
\usepackage{tikz}

\begin{document}


\pagenumbering{gobble}

\title{Single chain differential evolution Monte-Carlo for self-tuning Bayesian inference}
\author{Willem Bonnaff\'e$^1$}
\date{}
\maketitle

\begin{center}
\vspace{-0.5cm}
1. Big Data Institute, University of Oxford, Old Road Campus, Oxford OX3 7LF

\vspace{0.5cm}
\textbf{Abstract}
\end{center}

1. Bayesian inference is difficult because it often requires time consuming tuning of samplers.
Differential evolution Monte-Carlo (DEMC) is a self-tuning multi-chain sampling approach which requires minimal input from the operator as samples are obtained by taking the difference of the current position of multiple randomly selected chains.
However, this can also make DEMC more computationally intensive than single chain samplers.
2. We provide a single-chain adaptation of the DEMC algorithm by taking samples according to the difference in previous states of the chain, rather than the current state of multiple chains. 
This minimises computational costs by requiring only one posterior evaluation per step, while retaining the self-adaptive property of DEMC.
We test the algorithm by sampling a bivariate normal distribution and by estimating the posterior distribution of parameters of an ODE model fitted to an artificial prey-predator time series.
In both cases we compare the quality of DEMC generated samples to those obtained by a standard adaptive Markov chain Monte-Carlo sampler (AMC).
3. In both case studies, DEMC is as accurate as AMC in estimating posterior distributions, while being an order of magnitude faster  due to simpler computations.
DEMC also provides a higher effective samples size than AMC, and lower initial samples autocorrelations.
4. 
Its low computational cost and self-adaptive property make single chain DEMC particularly suitable for fitting models that are costly to evaluate, such as ODE models.
The simplicity of the algorithm also makes it easy to implement in base R, hence offering a simple alternative to STAN.


\textbf{Emails:}
willem.bonnaffe@nds.ox.ac.uk;

%


\newpage
\pagenumbering{arabic}
\setcounter{page}{1}
\setstretch{1.5}

\section{Introduction}

Bayesian inference differs from frequentist inference in that it provides parameter distributions instead of point estimates, which allow for a thorough and intuitive quantification of uncertainty.
This also makes Bayesian models more computationally intensive to fit, as they require numerical sampling schemes to estimate the posterior density distribution of model parameters.
However, modern computers have greatly reduced the computational costs of fitting Bayesian models.
As a result, Bayesian inference has gained popularity in ecology over the past decade.

Gradient-free Markov chain Monte-Carlo (MCMC) is the main family of methods for estimating parameter distributions.
These methods do not rely on the calculation of the derivative of the posterior distribution with respect to model parameters.
The chain moves in the parameter space according to a jump kernel, also known as proposal distribution (\cite{Hobbs2015}).
In most cases, the proposal distribution is defined as a multivariate Gaussian distribution, which handles covariations between parameters (\cite{Hobbs2015}).
The new parameter vector is accepted with a probability equal to the Metropolis-Hastings (MH) ratio, namely the ratio of the log posterior density of the proposed and current parameter vectors (\cite{Hobbs2015}).

Gradient-free MCMC implementations have evolved considerably over the past two decades.
Adaptive Monte-Carlo (AMC) introduced adaptation of the jump kernel to better match the shape of the target distribution (\cite{Haario2001}).
In the case of a Gaussian jump kernel, this amounts to updating the variance-covariance matrix with previous samples.
Later implementations introduced delayed rejections, where rejected samples are temporarily added to the chains, as a way to encourage further explorations of the parameter space (\cite{Haario2006}).
Several approaches were also developed to approximate the target distribution, either with PCA (\cite{Kennedy2015}), exponential family functions (\cite{Strathmann2015}), linear, quadratic, or Gaussian processes (\cite{Conrad2016}), in order to move more efficiently in the parameter space and reduce computational costs.
More recently, the no-U-turn sampler (NUTS), has gained popularity among ecologists, due to its capacity to handle high dimensional and hierarchical models (\cite{Monnahan2018}).
Yet, these algorithms are not equally well-suited to solve different ecological problems (\cite{Ponisio2019}), which often requires a trial and error search for a suitable implementation. 

A further limitation of these methods is that the quality of the sampling is contingent on an adequate tuning of the hyperparameters of the jump kernel, which is a difficult and time consuming task.
Ter Braak et al. developed a self-tuning approach, differential evolution Monte-Carlo (DEMC, \cite{TerBraak2006}), also known as DREAM (differential evolution adaptive Monte-Carlo, \cite{TerBraak2008}), by combining differential evolution (DE), a powerful genetic algorithm for global optimisation, and MCMC sampling.
Multiple chains explore the parameter space by choosing new samples via a linear combination of the current state of the other chains (\cite{TerBraak2006}).
To improve chain mixing, Ter Braak et al. further allowed new samples to be taken according to previous samples, in addition to the current state of the chains (\cite{TerBraak2008}).
The great strength of the approach is that very little input from the operator is required to perform the inference, as the algorithm itself tunes the jump kernel (\cite{TerBraak2006}).

Though DEMC is computationally competitive compared to MCMC, the need for multiple chains still increases computational cost.
We hence propose a single chain adaptation of the DEMC algorithm.
Using a single chain means that we only need to evaluate the posterior once per step, which makes the approach computationally inexpensive.
The shape of the posterior is learned by DE from previous states of the chain, with no gradient nor tuning required from the operator.
This makes the approach especially suitable for sampling costly mathematically intractable posterior distributions, such as that of ODE model parameters.

In spite of successful implementations in other fields (e.g. \cite{Turner2012,Turner2013,Shockley2018}), DEMC remains seldom used by ecologists (but see \cite{Lu2017}).
STAN and JAGS have increasingly been used to perform Bayesian inference of ecological models (\cite{Monnahan2018}).
Yet, they require a specific language, which makes it hard to customise models and interface with R (\cite{Monnahan2018}).
DEMC is a simple and robust sampling algorithm which can be implemented in only a few lines of code in base R, hence easing Bayesian inference.
Therefore, we believe it useful to reiterate the main strengths of DEMC in a context that may appeal to ecologists.

In this paper, we first introduce a single chain adaptation of the DEMC algorithm.
We provide a self-standing and ready to use implementation of the algorithm in R, Rcpp, and python.
In a first case study, we assess how well the sampler estimates the density of a bivariate normal distribution, for which true samples are known.
In a second case study, we provide a typical example of population dynamics analysis, by fitting an ODE model to prey-predator oscillations generated by the Lotka-Volterra system.
In both cases studies, we benchmark the approach against a standard AMC sampler.
We show that single chain DEMC is faster, and has a larger effective sample size (ESS), than AMC.
Our work establishes that single chain DEMC is a simple, fast, and robust sampling algorithm, which provides ecologists who wish to perform inference in base R with a simple, self-standing, alternative to STAN, or JAGS.

\section{Method}

\textbf{Population DEMC}

Differential evolution Monte-Carlo (DEMC) is an ensemble sampling scheme (\cite{TerBraak2006}). 
The ensemble consists of a population of chains that interact with each other to explore the target distribution. 
In DEMC, the steps are performed by differential evolution (DE), where a given chain uses the position of other randomly selected chains to move more efficiently than a random walker. 
More specifically, the step of a chain is proportional to the difference between the position of two other randomly selected chains in the parameter space, plus an error term.  
The error term encourages the chain to jump to unexplored areas of the parameter space.
DEMC hence does not require the specification and tuning of a proposal distribution, as the shape of the target is implicitly approximated by DE the jump kernel.
This relieves the operator from the need to tune the algorithm.
However, DEMC requires at least 3 chains, and preferably more (\cite{TerBraak2008}), which can make it more computationally intensive than simple adaptive MCMC.

\textbf{Single chain DEMC}

We propose here a single chain adaptation of the population DEMC algorithm.
The motivation behind this is that a single chain algorithm requires only a single evaluation of the posterior per step, which is useful for computationally intensive problems.
In addition, this makes the algorithm extremely simple, and hence easy to implement in any programming language.
The main difference between the population algorithm and the single chain algorithm is that instead of using other chains to perform a jump, we use previous states of the chain:

\vspace{-0.5cm}
\begin{equation}
\theta_p = \theta^{(k)} + \gamma \left( \theta^{(u)} - \theta^{(v)} \right) + \varepsilon,
\end{equation}

where $\theta_p$ is the proposed state, $\theta^{(k)}$ the current state, $\theta^{(u)}$ and $\theta^{(v)}$ are the $u^{th}$ and $v^{th}$ past states of the chain, and $\gamma$ and $\varepsilon$ are tuning parameters. 
As with the population algorithm, $\gamma = 2.38/\sqrt{2d}$ and $\varepsilon \sim \mathcal{U}(-\delta,\delta)$, where $\delta = 0.001$ should yield optimal results (\cite{TerBraak2006}).

\textbf{Algorithm}

The following algorithm can be used to sample the target distribution $\pi(.)$:

\begin{figure}[H]
\begin{center}
\begin{minipage}{1\linewidth}
\begin{itemize}
\item Initialisation $k=0$
\begin{itemize}
\item Set $\theta^{(0)}$ such that $\pi(\theta^{(0)})>0$
\item Set $\gamma = 2.38/\sqrt{2d}$ and $\delta = 0.001$
\end{itemize}
\item For each iteration $0 < k \leq k_{max}$:
\begin{itemize}
\item Draw $\theta^{(u)}$ and $\theta^{(v)}$ with $u \sim \mathcal{U}(0,k-1)$ and $v \sim \mathcal{U}(0,k-1)$ 
\item Draw $\varepsilon \sim \mathcal{U}(-\delta,\delta)$
\item Compute $\theta_p = \theta^{(k)} + \gamma \left( \theta^{(u)} - \theta^{(v)} \right) + \varepsilon$
\item Draw $q \sim \mathcal{U}(0,1)$
\item If $\pi(\theta_p)/\pi(\theta^{(k)}) > q$ then set $\theta^{(k+1)} = \theta_p$ 
\item Else set $\theta^{(k+1)} = \theta^{(k)}$
\end{itemize}
\end{itemize}
\end{minipage}
\end{center}
\end{figure}

\textbf{Detailed balance}

It is necessary to show that single chain DEMC retains the main properties of Monte-Carlo (MC) methods, namely that detailed balance and ergodicity are preserved.
Given the similarity of the population and single chain DEMC algorithms, we can use the proofs used by Ter Braak in his original paper (\cite{TerBraak2006}).

Detailed balance can be proven by showing that the transition kernel is symmetrical, which implies that each step is reversible. 
The probability density of the proposed value conditional on the current and past states of the chain is

\vspace{-0.5cm}
\begin{equation}
q(\theta_p | \theta^{(k)}, \theta^{(u)}, \theta^{(v)}) = f_U (\theta_p | \theta^{(k)} + \gamma ( \theta^{(u)} - \theta^{(v)} ), \delta ),
\end{equation}

where $q(\theta_p | .)$ is the transition kernel to $\theta_p$ given $\theta^{(k)}$, $\theta^{(u)}$, $\theta^{(v)}$, and $f_U$ the probability density function of a uniform distribution centred on $\theta^{(k)} + \gamma ( \theta^{(u)} - \theta^{(v)} )$ with spread $\delta$.

The total probability of jumping from the current to the proposed value is obtained by summing the transition kernel over all $k-1$ past states of the chain

\vspace{-0.5cm}
\begin{equation}
q(\theta_p | \theta^{(k)}) = \sum_{u}^{k-1} \sum_{v \ne u}^{k-1} f_U (\theta_p | \theta^{(k)} + \gamma ( \theta^{(u)} - \theta^{(v)} ), \delta ).
\end{equation}

Similarly the total probability of the reverse jump is 

\vspace{-0.5cm}
\begin{equation} \begin{aligned}
q(\theta^{(k)} | \theta_p) &= \sum_{u}^{k-1} \sum_{v \ne u}^{k-1} f_U (\theta^{(k)} | \theta_{p} - \gamma ( \theta^{(u)} - \theta^{(v)} ), \delta ). \\
\end{aligned} \end{equation}

By noting that setting $u=v$ and $v=u$ yields a mathematically identical transition kernel we show that $q(\theta_p | \theta^{(k)}) = q(\theta^{(k)} | \theta_p )$ which satisfies detailed balance.

\textbf{Ergodicity}

As with any adaptive schemes, ergodicity is more difficult to establish given that the proposal distribution changes with each iteration (\cite{Roberts2007}).
Ergodicity can be demonstrated by showing that the chain is not transient, aperiodic, irreducible (\cite{TerBraak2006}).
Nontransiency is guaranteed if the algorithm converges to a stationary distribution, which requires a stationary proposal distribution.
For adaptive schemes this condition is achieved by progressively reducing the adaptation (\cite{Haario2001}).
In practice this is done by sampling from an increasing part of the history of the chain (e.g. by taking samples from iteration $k/2$ to $k$, $k$ being the current iteration, \cite{Haario2001}).
Aperiodicity requires that the chain can return to its current state in a finite amount of steps, while irreducibility requires that any state can be reached within a finite amount of steps.
Both conditions are guaranteed here due to 
the error term $\varepsilon$, which generates a random walk behaviour, thus theoretically allowing for the exploration of the entire parameter space (\cite{TerBraak2006}).

\section{Case studies}

In order to assess the reliability and efficiency of our single chain DEMC algorithm we perform two tests.
The first test consists in estimating the density of a bivariate correlated normal distribution.
The second test is to recover the probability density of the parameters of a differential equation system.
In both tests, the algorithm is compared to a standard adaptive Metropolis-Hastings algorithm (AMC).

\textbf{Bivariate normal distribution}

We test the capacity of our algorithm to recover the probability density of a highly correlated bivariate normal distribution. 
The target distribution is given by

\vspace{-0.5cm}
\begin{equation}
\pi(\theta) = (2\pi)^{-1} det(\Sigma)^{-\frac{1}{2}} \exp \left( - \frac{1}{2} \theta^{T} \Sigma^{-1} \theta \right),
\end{equation}

where $\pi(\theta)$ refers to the posterior probability density of the parameter vector $\theta= [\theta_1, \theta_2]^T$, $\Sigma$ is the covariance matrix, the diagonal elements of which are $\sigma_1 = \sigma_2 = 1$ and the correlation coefficient is set to $\rho=0.99$.

We run three AMC chains of 10000 iterations with a Gaussian transition kernel. 
The scale parameter is set to $\alpha = 2.38/\sqrt{4}$ and the covariance matrix of the transition kernel is initially set to a two dimensional identity matrix $\Sigma_p=I_2$. 
Initially, 250 samples are taken without adaptation. 
A further 10000 samples are taken with adaptation of the covariance matrix at every step using an increasing part of the chain (i.e. samples from $k/2$ to $k$, $k$ being the current iteration). 
We also run three DEMC chains of 10000 iterations. 
The scale parameter is set to $\gamma = 2.38/\sqrt{4}$ and the noise parameter is set to $\delta=0.001$. 
Both DEMC and AMC chains are initiated at random starting points $\theta_i^{(0)} \overset{iid}{\sim} \mathcal{N}(0,1),~~i=\{1,2\}$.

\textbf{Differential equation system}

We test the capacity of single chain DEMC to recover the probability density of parameters of a differential equation system. 
Here we use the Lotka-Voltera differential equation system, which consists in a set of two coupled differential equations

\vspace{-0.5cm}
\begin{equation}
\left\{ \begin{aligned} 
\frac{dX}{dt} &= \alpha X - \beta X Y \\
\frac{dY}{dt} &= \gamma X Y - \delta Y \\
\end{aligned} \right.
\end{equation}

where the dynamics of the system $dZ/dt = [dX/dt, dY/dt ]^T$ are governed by the state vector is $Z=[X, Y ]^T$, $Z\in \mathbb{R}_+^{2}$ and the parameter vector is $\theta = [ \alpha, \beta, \gamma, \delta ]^T$, $\theta \in \mathbb{R}_+^{4}$.

We use this system to simulate an artificial time series that we fit the model to. 
To do this we set $\theta_{true} = [ \alpha = 1, \beta = 0.1, \gamma= 0.1, \delta = 1 ]^T$ and integrate the system on $t \in [0, 10]$. 
Then we simply sample the predictions of the model every time step $\Delta t = 1$. 

We define a simple Bayesian model to fit the differential equation system

\vspace{-0.5cm}
\begin{equation}
\pi(\theta | Z) \propto \prod_k p(Z_k | \bar{Z}(k |\theta)) p(\theta),
\end{equation}

where $Z_k$ is the observed state and $\bar{Z}(t|\theta) = Z_0 + \int_0^{k} d\bar{Z} $ the expected state of the system at the observed time step $k = \{0, \hdots, 10\}$. 
The likelihood is defined as a gamma distribution around the expected state of the system $Z_k \overset{iid}{\sim} \Gamma(\bar{Z}(k|\theta),1),~\forall~k\in{0,\hdots,10}$. 
The prior distribution of the parameter vector is defined as a multivariate uniform distribution $\theta \sim \mathcal{U}_4(0,2)$.

We run three AMC chains of 10000 iterations. 
The scale parameter is set to $\alpha = 2.38/\sqrt{8}$ the covariance matrix of the transition kernel is initiated as a multivariate identity matrix $\Sigma_p = I_4$. 
Initially, 250 samples are taken without adaptation. 
A further 10000 samples are taken with adaptation of the covariance matrix at every step using an increasing part of the chain (i.e. samples from $k/2$ to $k$, $k$ being the current iteration). 
We also run three DEMC chains of 10000 iterations. 
As before the scale parameter is set to $\gamma = 2.38/\sqrt{8}$ and the noise parameter is set to $\delta = 0.001$
All chains are initiated at points randomly sampled from the prior distribution. 

\textbf{Analysis and comparison of algorithms}

We compare the AMC and DEMC algorithms on three levels.
First, we observe the trace of the chains to assess the rate of convergence of the algorithms.
Second, we compute the autocorrelation in the chain samples up to the $50^{th}$ order and effective sample size (\cite{Hobbs2015}).
The first 1000 iterations of the chains are removed to eliminate the transient states.
Finally, we compare the estimated quantiles of the stationary probability distributions obtained from the two algorithms. 
In this case the chains are further thinned down to 1000 independent samples. 
Results are presented below.

\newpage
\section{Results}

\textbf{Bivariate normal distribution}

Single chain DEMC and AMC converge to virtually identical posterior distributions, which approximate well the shape of the bivariate normal distribution (Figure 1). 
In doing so, our single chain DEMC algorithm is about an order of magnitude faster than AMC (Table 1).
The autocorrelation in the samples of DEMC drops faster than that of AMC which remains slightly above 0 for longer (Figure 2). 
Samples estimated by single chain DEMC match up with both those generated by AMC and by the true distribution (Figure 3).

\vspace{0.5cm}
\begin{table}[H]
\caption{
    \textbf{Summary results of sampling the bivariate normal distribution with AMC and DEMC.} 
    The Dim column corresponds to the dimension of the bivariate normal, the MaP, mean and sd are the maximum a posteriori, mean and standard deviation estimated from the samples, $\hat{r}$ is the convergence statistic (as in \cite{Hobbs2015}), and ESS denotes the effective sample size. 
    The time column corresponds to the time in seconds required to run 3 chains of 10000 iterations.
}
\begin{center}
\begin{tabular}{cccccccc}
\hline
\\
Algorithm & Dim. & MaP & mean & sd & $\hat{r}$ & Time & ESS \\
\\
\hline
\\
AMC   & $\theta_1$ & -0.0042 & -0.0105 & 1.0001 & 1.0024 & 77.163 & 186.4  \\
      & $\theta_2$ & -0.0087 & -0.0097 & 1      & 1.0022 & -      & 183.56 \\
\\
DE-MC & $\theta_1$ & -0.0328 & -0.0013 & 0.9893 & 1.0006 & 4.007  & 223.96 \\
      & $\theta_2$ & -0.0332 & -0.006  & 0.9928 & 1.0005 & -      & 222.47 \\
\end{tabular}
\end{center}
\end{table}

\begin{figure}[H]
\begin{center}
\includegraphics[page=1, width=1\linewidth]{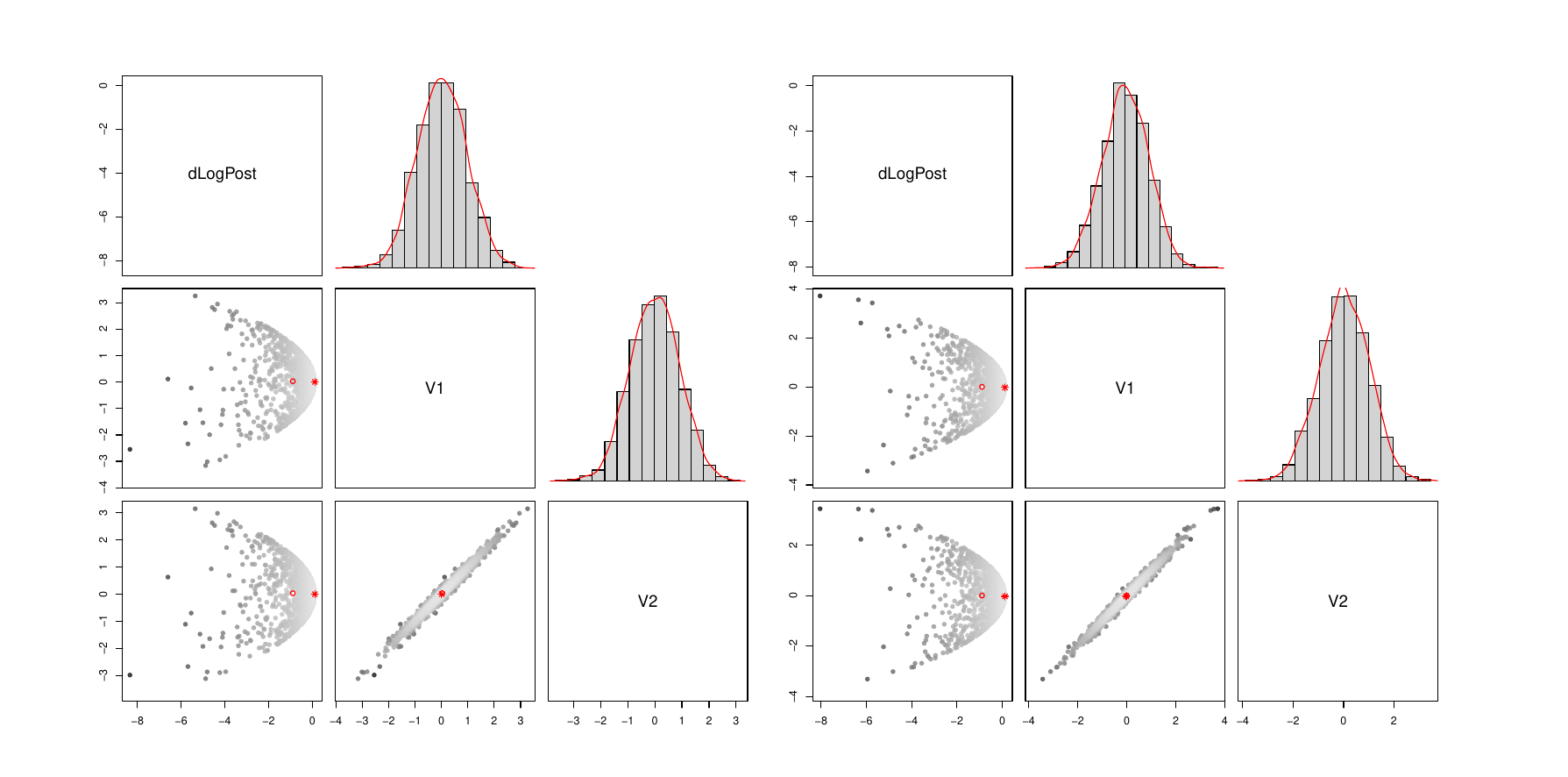}
    \caption{
        \textbf{Estimated posterior distributions of the parameters of the bivariate normal model.} 
        Comparison of the estimated posterior distributions of parameters $\theta_1$ and $\theta_2$ of the bivariate normal model, obtained by AMC (left) and DEMC (right). 
        dLogPost corresponds to the log-density of the bivariate normal distribution.
        V1 and V2 correspond to the first and second parameter, $\theta_1$ and $\theta_2$, respectively. 
        The shade of grey, from dark to light, is proportional to the posterior density.
        The histograms correspond to the marginal probability density of the parameters.
        Final samples are obtained by burning and thinning the chains to retain 1000 samples.
    }\label{fig:post_MVN}
\end{center}
\end{figure}

\newpage
\begin{figure}[H]
\begin{center}
\includegraphics[page=2, width=1\linewidth]{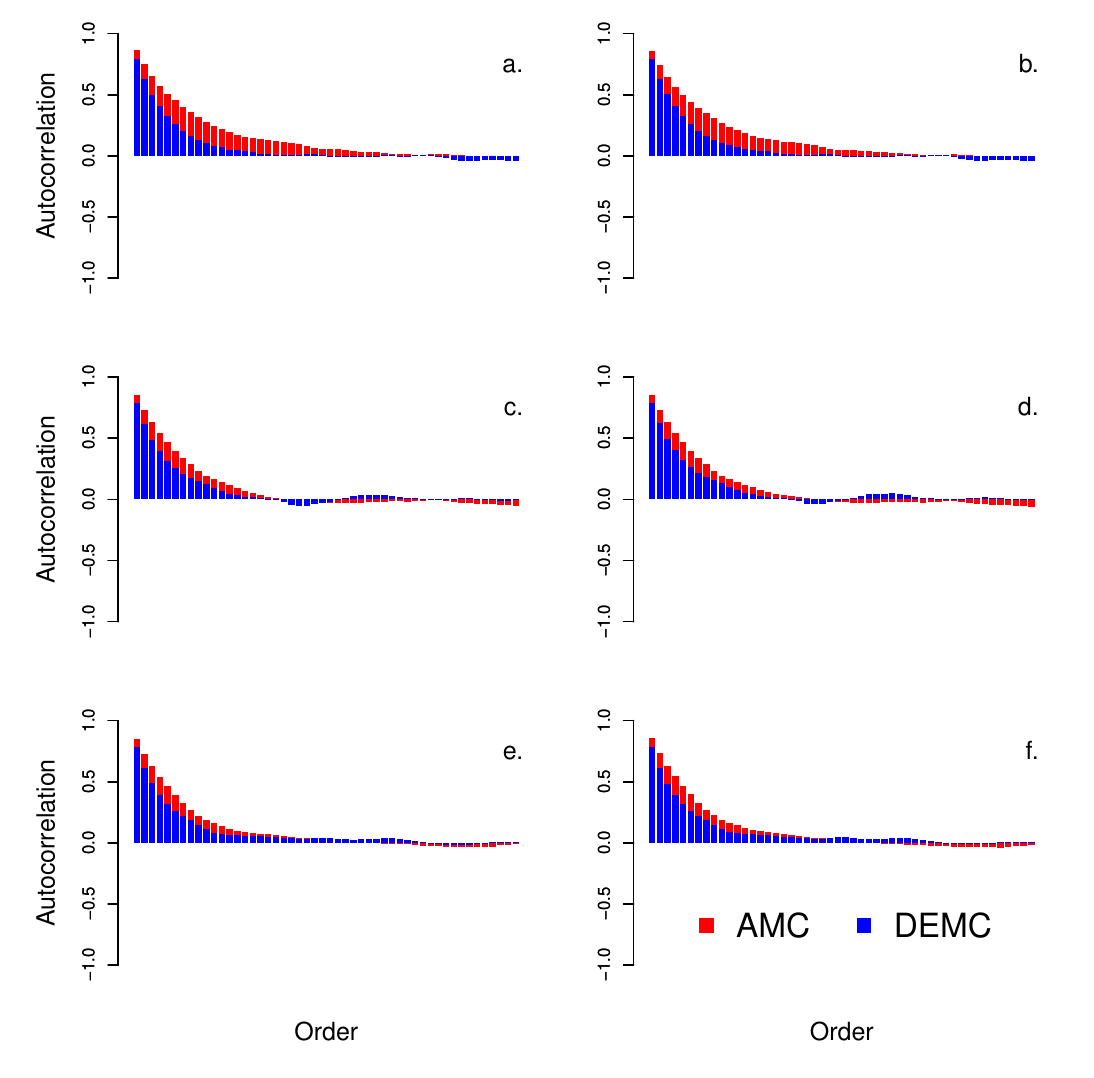}
    \caption{
        \textbf{Sample autocorrelations of the bivariate normal model.} 
        Comparison of autocorrelations of samples generated by AMC (red) and DEMC (blue). 
        Each row corresponds to one of three chains. 
        Each column corresponds to each parameter of the bivariate Gaussian.
        Graph a., c., and e., correspond to the three chains for parameter $\theta_1$, while graph b., d., and f. correspond to parameter $\theta_2$.
        Autocorrelations are calculated on the chains before thinning.
    } \label{fig:autocor_MVN}
\end{center}
\end{figure}

\newpage
\begin{figure}[H]
\begin{center}
\includegraphics[page=3, width=1\linewidth]{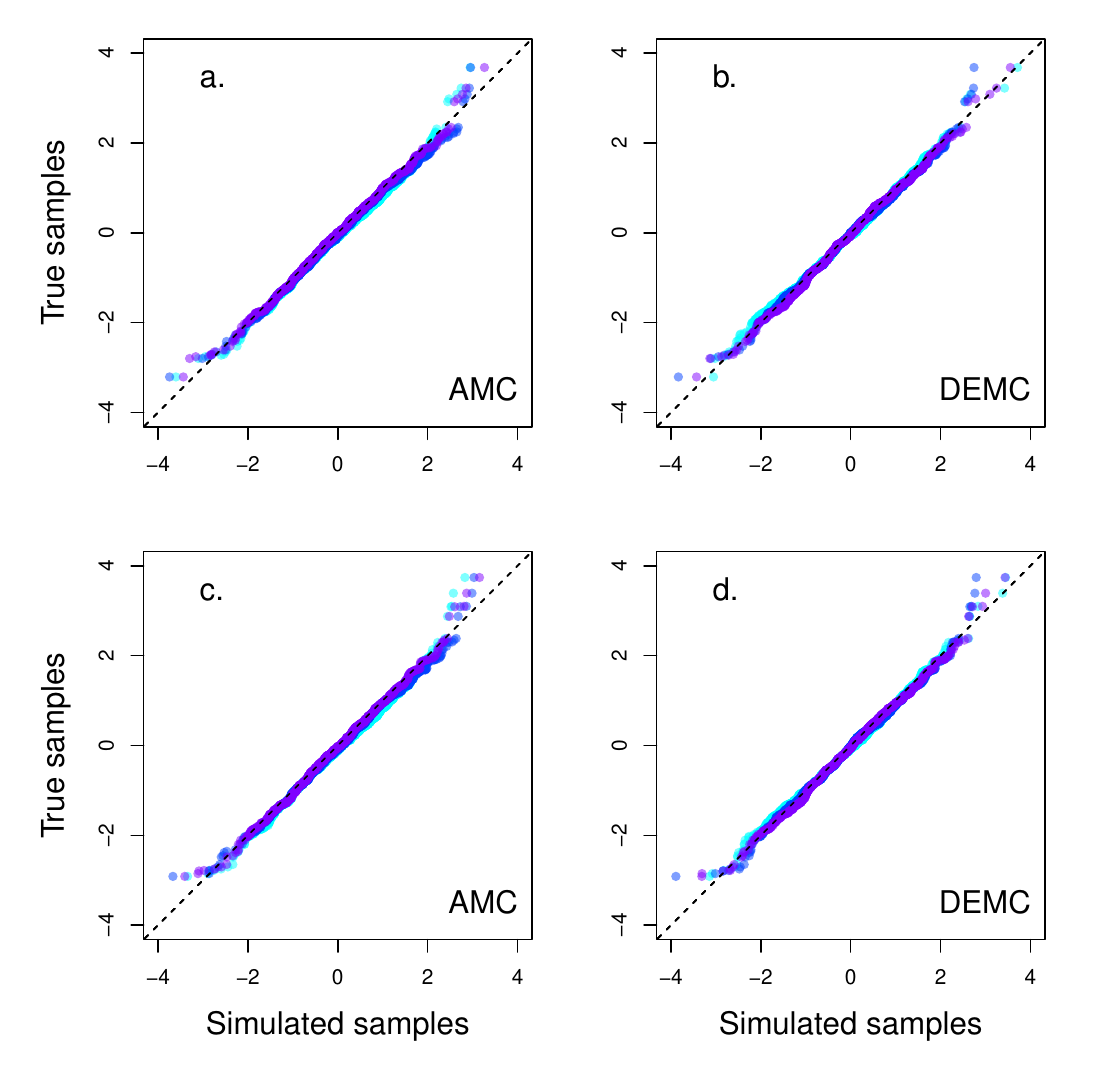}
    \caption{
        \textbf{Samples of the bivariate normal model.} 
        Comparison of true samples taken from the bivariate normal distribution and samples obtained by AMC (a. and c.) and DEMC (b. and d.) for parameter $\theta_1$ (a. and b.) and $\theta_2$ (c. and d.). 
        Different colours correspond to different chains.
    } \label{fig:samples_MVN}
\end{center}
\end{figure}
\newpage

\textbf{ODE system}

In the ODE case study, the chains of single chain DEMC also converge faster to the target distribution than AMC (Table 2).
The posterior distributions estimated by AMC and DEMC are virtually identical (Figure 4). 
The autocorrelation in the samples of single chain DEMC drops faster than  AMC (Figure 5). 
Finally, parameter samples obtained by AMC and DEMC are virtually identical (Figure 6).

\vspace{0.5cm}
\begin{table}[H]
\caption{
    \textbf{Summary results of fitting the ODE model with the AMC and DEMC.} 
    The Dim column corresponds to the parameters of the ODE model, the MaP, mean and sd are the maximum a posteriori, mean and standard deviation estimated from the samples, $\hat{r}$ is the convergence statistic (as in \cite{Hobbs2015}), and ESS denotes the effective sample size. 
    The time column corresponds to the time in seconds required to run 3 chains of 10000 iterations.
}
\begin{center}
\begin{tabular}{cccccccc}
\hline
\\
Algorithm & Dim. & MaP & mean & sd & $\hat{r}$ & Time & ESS \\
\\
\hline
\\
AMC & $\alpha$ & 1.0029 & 0.999  & 0.0353 & 1.002  & 389.223 & 151.31 \\
    & $\beta$  & 0.1007 & 0.1014 & 0.0105 & 1.0006 & -       & 154.55 \\
    & $\gamma$ & 0.0984 & 0.1014 & 0.0134 & 1.0022 & -       & 147.44 \\
    & $\delta$ & 0.9688 & 0.975  & 0.0891 & 1.0032 & -       & 144.15 \\
\\
DE-MC & $\alpha$ & 1.0078 & 0.9972 & 0.036  & 1.0007 & 115.251 & 182.21 \\
      & $\beta$  & 0.1017 & 0.1011 & 0.0109 & 1.0006 & -       & 172.34 \\
      & $\gamma$ & 0.097  & 0.1019 & 0.0136 & 1.0009 & -       & 178.13 \\
      & $\delta$ & 0.9608 & 0.9743 & 0.0864 & 1      & -       & 183.54 \\
\end{tabular}
\end{center}
\end{table}

\begin{figure}[H]
\begin{center}
\includegraphics[page=4, width=1\linewidth]{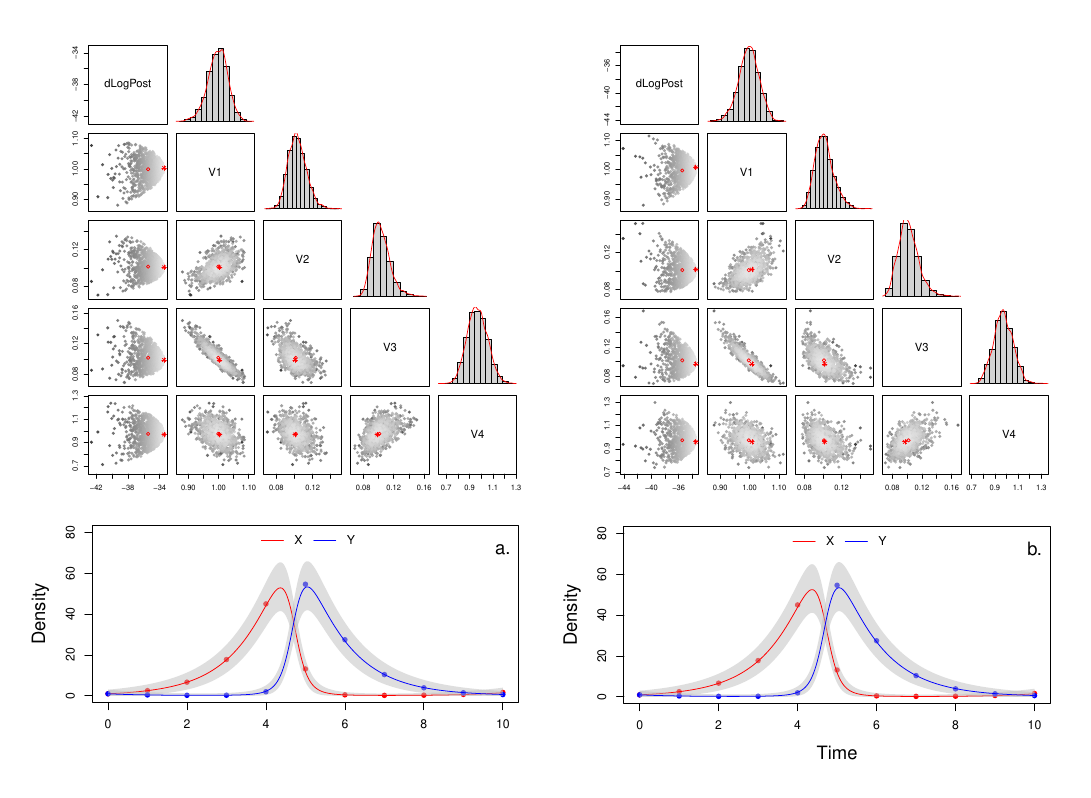}
    \caption{
        \textbf{Estimated posterior distributions of the parameters of the ODE model.} 
        Comparison of the estimated posterior distributions of parameters $\alpha$, $\beta$, $\gamma$, and $\delta$ of the Lotka-Volterra ODE system, obtained by AMC (left) and DEMC (right). 
        dLogPost corresponds to the log-density of the bivariate normal distribution.
        V1, V2, V3, and V4 correspond to the ODE parameters, $\alpha$, $\beta$, $\gamma$, and $\delta$, respectively. 
        The shade of grey, from dark to light, is proportional to the posterior density.
        The histograms correspond to the marginal probability density of the parameters.
        Final samples are obtained by burning and thinning the chains to retain 1000 samples.
        The fit of the ODE models to the artificial time series is shown in graph a. (AMC) and b. (DEMC).
        Solid lines, either red for the prey, X, or blue for the predator, Y, correspond to the best fit, and the shaded area is the 90\% confidence interval on predictions.
    }\label{fig:post_ODE}
\end{center}
\end{figure}

\newpage
\begin{figure}[H]
\begin{center}
\includegraphics[page=5, width=1\linewidth]{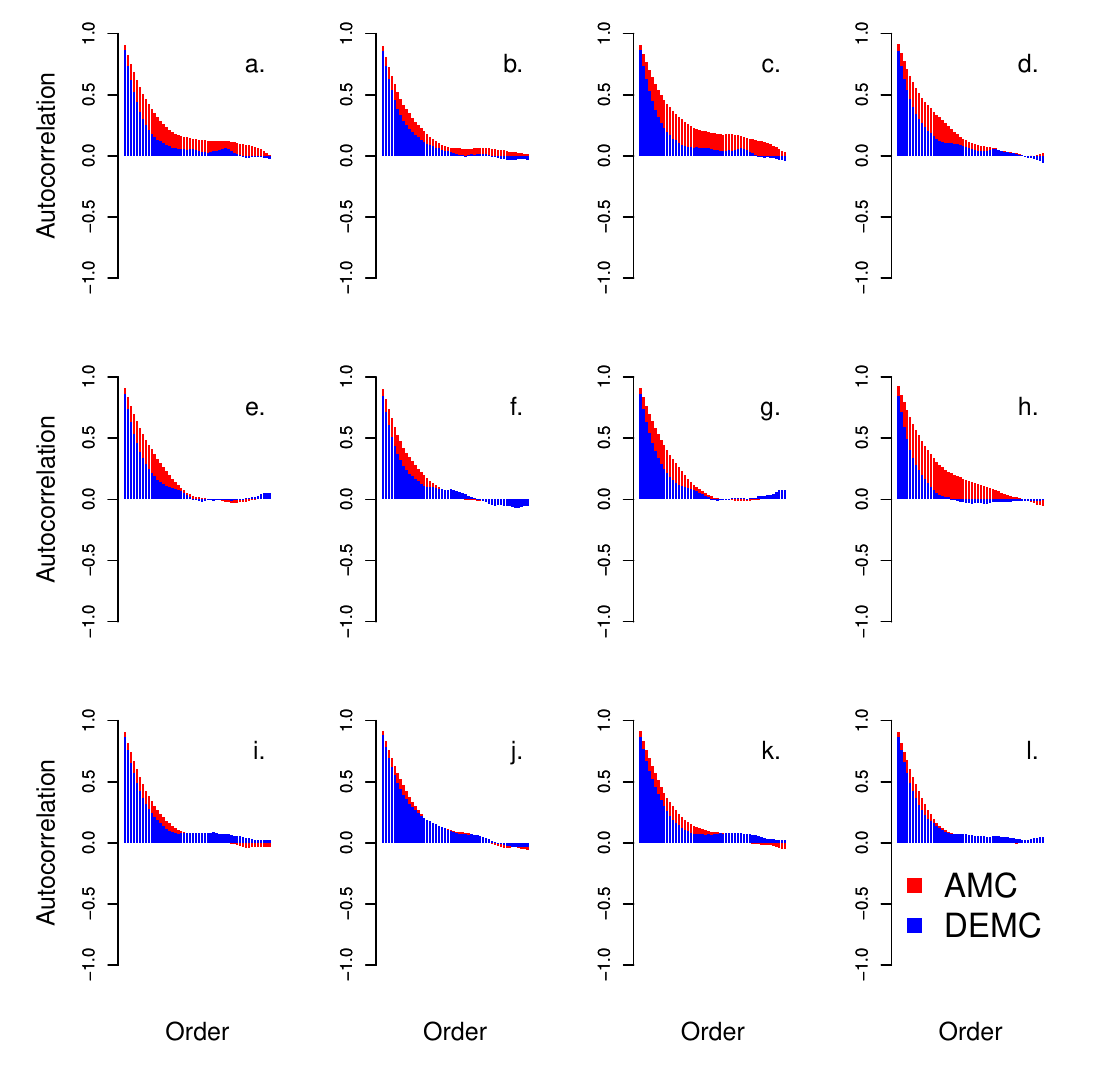}
    \caption{
        \textbf{Sample autocorrelations of the ODE model.} 
        Comparison of autocorrelations of samples generated by AMC (red) and DEMC (blue). 
        Each row corresponds to one of three chains. 
        Each column corresponds to each parameter of the bivariate Gaussian.
        Graph a., c., and e., correspond to the three chains for parameter $\theta_1$, while graph b., d., and f. correspond to parameter $\theta_2$.
        Autocorrelations are calculated on the chains before thinning.
    } \label{fig:autocor_ODE}

\end{center}
\end{figure}

\newpage
\begin{figure}[H]
\begin{center}
\includegraphics[page=6, width=1\linewidth]{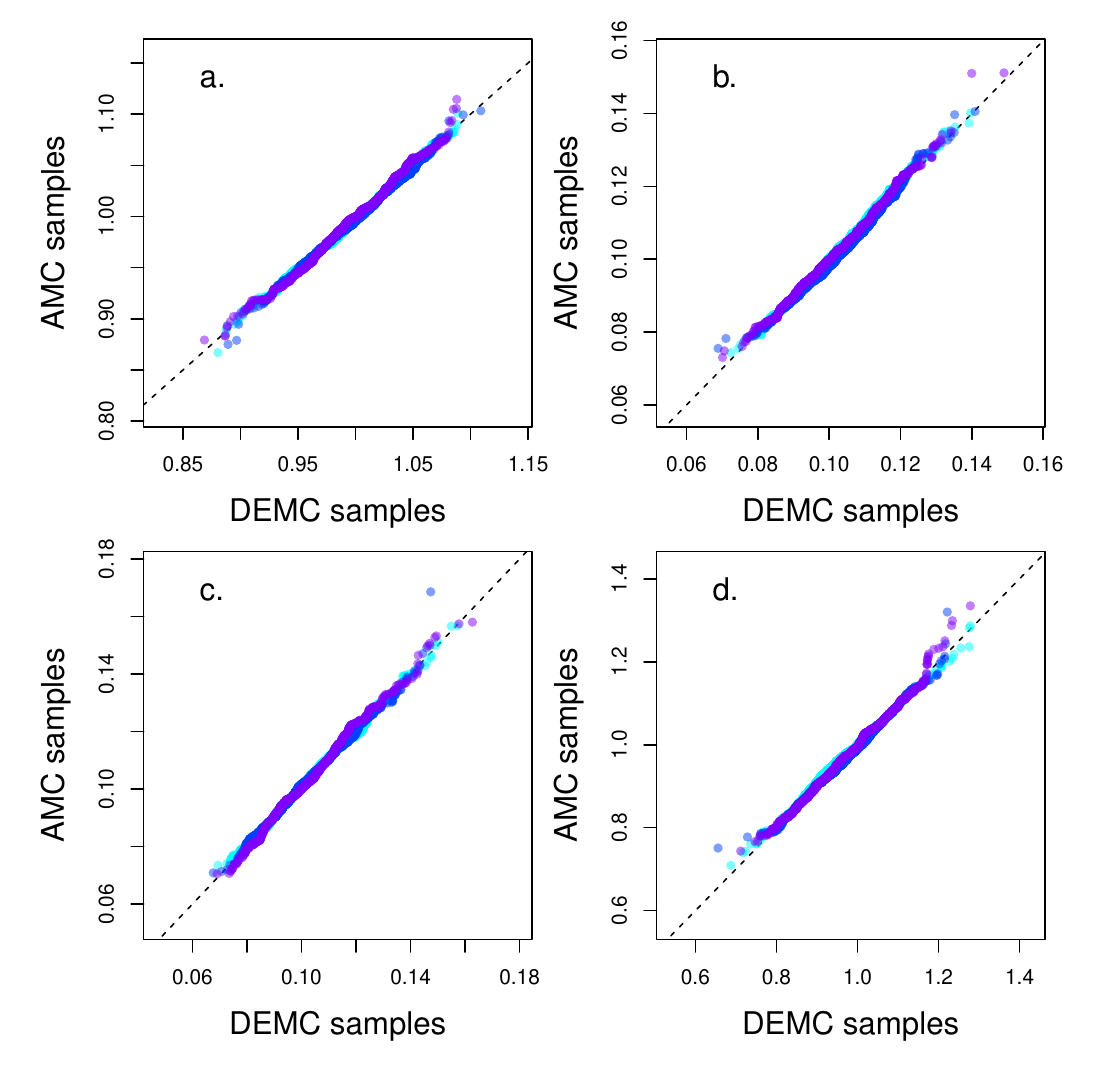}
    \caption{\textbf{Samples of the ODE model.} 
        Comparison of samples obtained by AMC (y axis) and DEMC (x axis) for each of the four parameters of the ODE system, $\alpha$ (a.), $\beta$ (b.), $\gamma$ (c.), $\delta$ (d.). 
        Different colours correspond to different chains.
    } \label{fig:samples_ODE}

\end{center}
\end{figure}

\section{Discussion}

Bayesian inference is challenging because of the need for sampling the posterior distributions of model parameters.
Differential evolution Monte-Carlo (DEMC) is a self-adaptive sampling method, which relieves the operator from the need of tuning the sampler.
The method however requires an ensemble of chains, which can make more computationally intensive to use than single chain alternatives.
Here we provide a single chain adaptation of DEMC, making it computationally inexpensive and simple to implement.
We show that our method is accurate by sampling a bivariate normal distribution where true samples are known.
We also show that the method is accurate in fitting ODE models to prey-predator time series data.
In both case studies we compare results of single chain DEMC to a standard adaptive Monte-Carlo (AMC).
We find that our method is quantitatively identical to AMC while being much faster.
Single chain DEMC hence provides a simple, accurate, and robust alternative to performing Bayesian inference in STAN and JAGs.

\textbf{Comparison of single chain DEMC to AMC, DREAM, and HMC}

We find that parameter estimation by single chain DEMC and AMC are identical, for both the bivariate normal model and ODE model.
However, single chain DEMC is much faster than AMC.
This is due to the fact that tuning the jump kernel of AMC involves computing a covariance matrix of previous samples (\cite{Haario2001}), which is computationally expensive.
In contrast, DEMC only requires taking the difference between two previous samples, which is inexpensive (\cite{TerBraak2006}).
In addition, sample autocorrelations of DEMC are lower than AMC, which translates into a higher effective sample size for DEMC.
This may be due to the fact that the jump kernel of DEMC has no predetermined functional form, hence making it better at approximating different target shapes than AMC.

The original implementation of DEMC (also known as DREAM) involves multiple chains (\cite{TerBraak2006}).
The new location of the chain in the parameter space is determined by taking the difference of the current location of the other chains, chosen at random.
Using multiple chains offers several advantages over a single chain implementation.
First, it reduces the risk of getting trapped in a local maximum.
Ter Braak also noted that chain mixing increased with the number of elements in the ensemble (\cite{TerBraak2008}).
However, computational cost also increases with the number of chains.
So that, even if single chain DEMC is potentially less powerful than multiple chain DEMC, it makes up for it by taking more samples per unit of time.

We choose not to compare HMC and DEMC because we think they serve different purposes.
HMC is only suitable for targets that are not too costly to evaluate and for which gradients are available (\cite{Strathmann2015}).
DEMC is more suitable for computationally expensive and mathematically intractable targets.
Regardless, even in cases where HMC may be more suitable, DEMC would be useful for fast prototyping the model and running pilot chains, due to its simplicity and the fact that it does not require gradients.

\textbf{Suitability of single chain DEMC for parameter estimation in ODE models}

Fitting ODE models to time series data is a powerful technique to identify the drivers of population dynamics (\cite{Demyanov2006,Rosenbaum2019,Adams2020,Bonnaffe2021a,Bonnaffe2022}).
For instance, ODE parameters can describe the birth, death rate of individuals, as well as the effect of competitors and predators in population growth.
Estimating ODE parameters is challenging because the predictions of ODE models are obtained from numerical simulation (e.g. \cite{Bonnaffe2021a, Bonnaffe2022}).
This makes it computationally expensive to evaluate the posterior distribution, and prevents the calculation of gradients.
Single chain DEMC is highly suitable for fitting these models.
This is because only one evaluation of the posterior is required per step, hence reducing computational costs as much as possible.
The self tuning property of single DEMC also alleviates the need for running pilot chains and tuning the algorithm, thus saving the operator time.
We demonstrate this in our second case study, where we fit an ODE model to a prey-predator time series generated by a Lotka-Volterra model.
Single chain DEMC provides accurate parameter estimates compared to ground truth, while being much faster than AMC.
Single chain DEMC hence takes the hassle away from the fitting, allowing the operator to focus on designing and comparing ODE models.

\textbf{Limits and prospects}

In our single chain DEMC algorithm, a new sample is taken by computing the difference between two previous samples selected at random.
This, in essence, is a linear interpolation/extrapolation of the previous states of the chain.
While we show that this performs well on correlated/nonlinear distributions, a nonlinear combination of previous states may perform better.
To do this, we could use nonlinear combinations of the previous states, by using polynomial functions, or even artificial neural networks (ANN).
One difficulty with this is that is requires extra tuning parameters to control the nonlinearity of the jump kernel.
However, those could be treated as ordinary model parameters and learned by differential evolution as the chain is running.
We view neural differential evolution Monte-Carlo as a particularly promising prospect given the powerful approximation capacity of ANNs.

Finally, we only considered a simple implementation of single chain DEMC in R, Rcpp, and python.
We did not exploit the potential for CPU and GPU parallelization of the algorithm.
This could result in order of magnitude faster inference.

\textbf{Conclusion}

In conclusion, we demonstrate that our single chain adaptation of the DEMC algorithm is easy to use and implement, computationally efficient, and improves sampling efficiency. 
The differential evolution step alleviates the need for trial and error tuning of the sampler, allowing the operator to focus on model design and comparison.
The simplicity of the algorithm makes it easy to implement in base R in a few lines of code, hence providing an easy alternative to more complicated Bayesian inference tools (e.g. STAN, JAGS).
The single chain nature of the algorithm makes it computationally inexpensive compared to multiple chain DEMC.
Overall, our algorithm is particularly suitable for Bayesian inference of computationally expensive and complicated models, such as ODE models, that require an algorithm that minimises model evaluations. 

\textbf{Statement of authorship}

Willem Bonnaff\'e designed the method, performed the analysis, wrote the manuscript;

\textbf{Acknowledgments}

We thank warmly Tim Coulson and the Ecological and Evolutionary Lab for comments and feedback on early versions of the work.
This work was supported by the Oxford-Oxitec scholarship and the NERC DTP.

\textbf{Data accessibility}

All data and code is available at https://github.com/WillemBonnaffe/DEMC.

\setstretch{1.25}
\printbibliography

@article{Haario2001,
   author = {Heikki Haario and Eero Saksman and Johanna Tamminen},
   doi = {10.2307/3318737},
   issn = {13507265},
   issue = {2},
   journal = {Bernoulli},
   pages = {223},
   title = {An Adaptive Metropolis Algorithm},
   volume = {7},
   year = {2001},
}

@article{TerBraak2006,
   author = {Cajo J. F. Ter Braak},
   doi = {10.1007/s11222-006-8769-1},
   issn = {0960-3174},
   issue = {3},
   journal = {Statistics and Computing},
   pages = {239-249},
   title = {A Markov Chain Monte Carlo version of the genetic algorithm Differential Evolution: easy Bayesian computing for real parameter spaces},
   volume = {16},
   year = {2006},
}

@article{Demyanov2006,
   author = {V. Demyanov and S. N. Wood and T. J. Kedwards},
   doi = {10.1111/j.1467-9876.2005.00527.x},
   issn = {00359254},
   issue = {1},
   journal = {Journal of the Royal Statistical Society. Series C: Applied Statistics},
   pages = {41-62},
   title = {Improving ecological impact assessment by statistical data synthesis using process-based models},
   volume = {55},
   year = {2006},
}

@article{Haario2006,
   author = {Heikki Haario and Marko Laine and Antonietta Mira and Eero Saksman},
   doi = {10.1007/s11222-006-9438-0},
   issn = {09603174},
   issue = {4},
   journal = {Statistics and Computing},
   month = {12},
   pages = {339-354},
   title = {DRAM: Efficient adaptive MCMC},
   volume = {16},
   year = {2006},
}

@article{TerBraak2008,
   author = {Cajo J.F. Ter Braak and Jasper A. Vrugt},
   doi = {10.1007/s11222-008-9104-9},
   issn = {09603174},
   issue = {4},
   journal = {Statistics and Computing},
   pages = {435-446},
   title = {Differential Evolution Markov Chain with snooker updater and fewer chains},
   volume = {18},
   year = {2008},
}

@article{Roberts2007,
   author = {Gareth Roberts and Jeffrey Rosenthal},
   issue = {2},
   journal = {Journal of Applied Probability},
   pages = {458-475},
   title = {Coupling and Ergodicity of Adaptive Markov Chain Monte Carlo Algorithms},
   volume = {44},
   year = {2012},
}

@article{Turner2012,
   author = {Brandon M. Turner and Per B. Sederberg},
   doi = {10.1016/j.jmp.2012.06.004},
   issn = {00222496},
   issue = {5},
   journal = {Journal of Mathematical Psychology},
   month = {10},
   pages = {375-385},
   title = {Approximate Bayesian computation with differential evolution},
   volume = {56},
   year = {2012},
}

@article{Turner2013,
   author = {Brandon M. Turner and Per B. Sederberg and Scott D. Brown and Mark Steyvers},
   doi = {10.1037/a0032222},
   issn = {1082989X},
   issue = {3},
   journal = {Psychological Methods},
   pages = {368-384},
   title = {A Method for efficiently sampling from distributions with correlated dimensions},
   volume = {18},
   year = {2013},
}

@book{Hobbs2015,
   author = {N Thompson Hobbs and Mevin B Hooten},
   city = {Princeton},
   doi = {doi:10.1515/9781400866557},
   isbn = {9781400866557},
   publisher = {Princeton University Press},
   title = {A Statistical Primer for Ecologists},
   url = {https://doi.org/10.1515/9781400866557},
   year = {2015},
}

@article{Strathmann2015,
   author = {Heiko Strathmann and Dino Sejdinovic and Samuel Livingstone and Zoltan Szabo and Arthur Gretton},
   journal = {arXiv},
   pages = {1-9},
   title = {Gradient-free Hamiltonian Monte Carlo with Efficient Kernel Exponential Families},
   url = {https://github.com/karlnapf/kernel_hmc},
   year = {2015},
}

@article{Kennedy2015,
   author = {David A. Kennedy and Vanja Dukic and Greg Dwyer},
   doi = {10.1007/s10651-014-0297-0},
   issn = {15733009},
   issue = {2},
   journal = {Environmental and Ecological Statistics},
   month = {6},
   pages = {247-274},
   publisher = {Kluwer Academic Publishers},
   title = {Combining principal component analysis with parameter line-searches to improve the efficacy of Metropolis–Hastings MCMC},
   volume = {22},
   year = {2015},
}

@article{Conrad2016,
   author = {Patrick R. Conrad and Youssef M. Marzouk and Natesh S. Pillai and Aaron Smith},
   doi = {10.1080/01621459.2015.1096787},
   issn = {1537274X},
   issue = {516},
   journal = {Journal of the American Statistical Association},
   month = {10},
   pages = {1591-1607},
   publisher = {American Statistical Association},
   title = {Accelerating Asymptotically Exact MCMC for Computationally Intensive Models via Local Approximations},
   volume = {111},
   year = {2016},
}

@article{Lu2017,
   author = {Dan Lu and Daniel Ricciuto and Anthony Walker and Cosmin Safta and William Munger},
   doi = {10.5194/bg-14-4295-2017},
   issn = {17264189},
   issue = {18},
   journal = {Biogeosciences},
   month = {9},
   pages = {4295-4314},
   publisher = {Copernicus GmbH},
   title = {Bayesian calibration of terrestrial ecosystem models: A study of advanced Markov chain Monte Carlo methods},
   volume = {14},
   year = {2017},
}

@article{Shockley2018,
   author = {Erin M. Shockley and Jasper A. Vrugt and Carlos F. Lopez},
   doi = {10.1093/bioinformatics/btx626},
   issn = {14602059},
   issue = {4},
   journal = {Bioinformatics},
   month = {2},
   pages = {695-697},
   pmid = {29028896},
   publisher = {Oxford University Press},
   title = {PyDREAM: High-dimensional parameter inference for biological models in python},
   volume = {34},
   year = {2018},
}

@article{Monnahan2018,
   author = {Cole C. Monnahan and Kasper Kristensen},
   doi = {10.1371/journal.pone.0197954},
   issn = {19326203},
   issue = {5},
   journal = {PLoS ONE},
   month = {5},
   pmid = {29795657},
   publisher = {Public Library of Science},
   title = {No-U-turn sampling for fast Bayesian inference in ADMB and TMB: Introducing the adnuts and tmbstan R packages},
   volume = {13},
   year = {2018},
}

@article{Rosenbaum2019,
   author = {Benjamin Rosenbaum and Michael Raatz and Guntram Weithoff and Gregor F. Fussmann and Ursula Gaedke},
   doi = {10.3389/fevo.2018.00234},
   issn = {2296701X},
   issue = {234},
   journal = {Frontiers in Ecology and Evolution},
   pages = {1-14},
   title = {Estimating parameters from multiple time series of population dynamics using bayesian inference},
   volume = {6},
   year = {2019},
}

@article{Adams2020,
   author = {Matthew P. Adams and Scott A. Sisson and Kate J. Helmstedt and Christopher M. Baker and Matthew H. Holden and Michaela Plein and Jacinta Holloway and Kerrie L. Mengersen and Eve McDonald-Madden},
   doi = {10.1111/ele.13465},
   issn = {14610248},
   issue = {4},
   journal = {Ecology Letters},
   month = {4},
   pages = {607-619},
   pmid = {31989772},
   publisher = {Blackwell Publishing Ltd},
   title = {Informing management decisions for ecological networks, using dynamic models calibrated to noisy time-series data},
   volume = {23},
   year = {2020},
}

@article{Ponisio2019,
   author = {Lauren C. Ponisio and Perry de Valpine and Nicholas Michaud and Daniel Turek},
   doi = {10.1002/ece3.6053},
   issn = {20457758},
   issue = {5},
   journal = {Ecology and Evolution},
   month = {3},
   pages = {2385-2416},
   publisher = {John Wiley and Sons Ltd},
   title = {One size does not fit all: Customizing MCMC methods for hierarchical models using NIMBLE},
   volume = {10},
   year = {2020},
}

@article{Bonnaffe2021a,
   author = {Willem Bonnaffé and Ben C. Sheldon and Tim Coulson},
   doi = {10.1111/2041-210x.13606},
   issn = {2041-210X},
   journal = {Methods in Ecology and Evolution},
   pages = {1-46},
   title = {Neural ordinary differential equations for ecological and evolutionary time series analysis},
   volume = {2},
   year = {2021},
}

@article{Bonnaffe2022,
   author = {Willem Bonnaffé and Tim Coulson},
   journal = {arXiv},
   month = {9},
   pages = {1-57},
   title = {Fast fitting of neural ordinary differential equations by Bayesian neural gradient matching to infer ecological interactions from time series data},
   url = {http://arxiv.org/abs/2209.06184},
   year = {2022},
}

%


\newpage
\section{Supplementary}
\appendix
\beginsupplement

\section{Traces of AMC chains fitted to the multivariate Gaussian}

\begin{figure}[H]
\begin{center}
\includegraphics[page=7, width=1\linewidth]{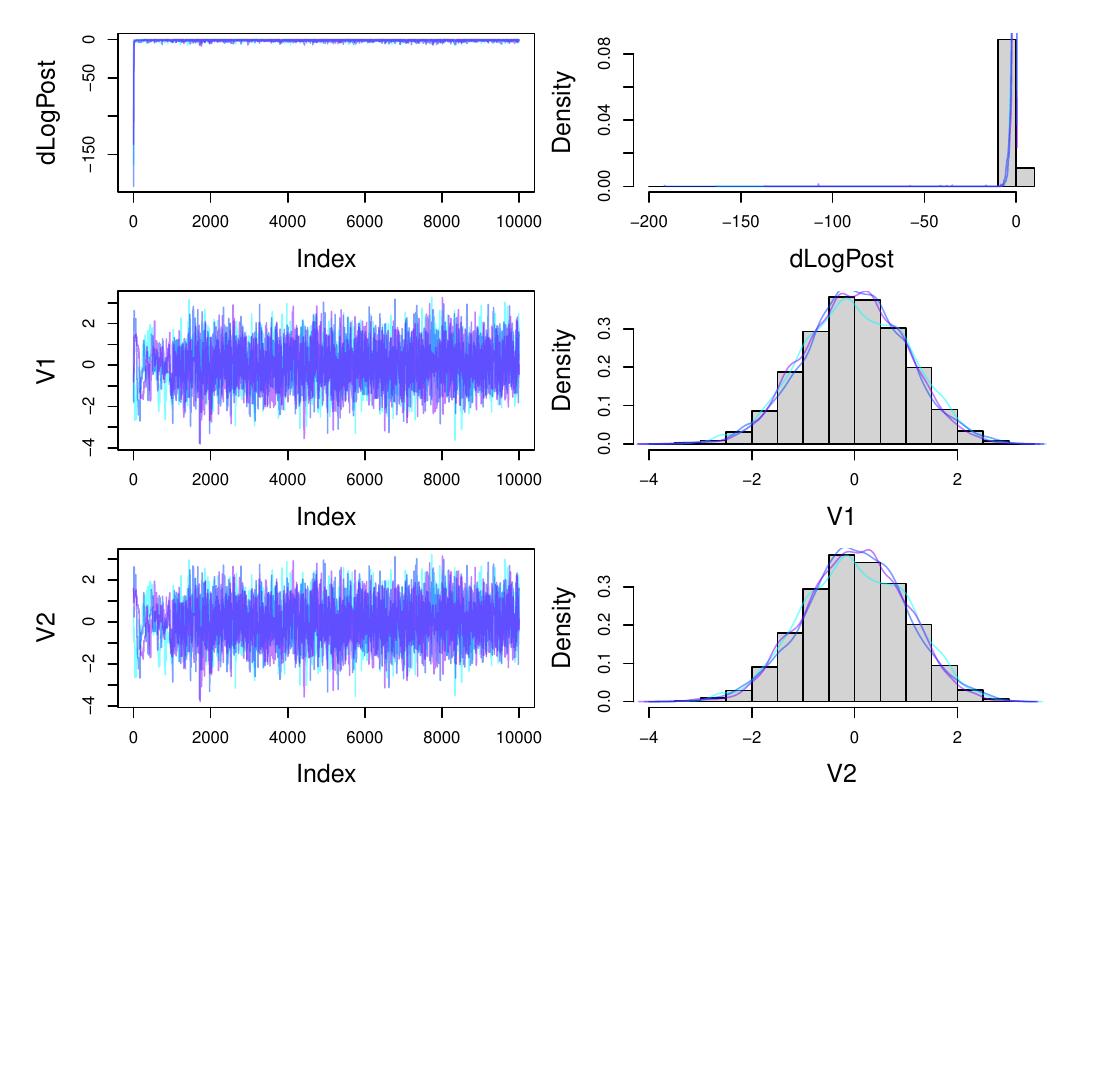}
\caption{\textbf{Trace of full AMC chains sampling the bivariate normal model.}} 
\end{center}
\end{figure}

\begin{figure}[H]
\begin{center}
\includegraphics[page=8, width=1\linewidth]{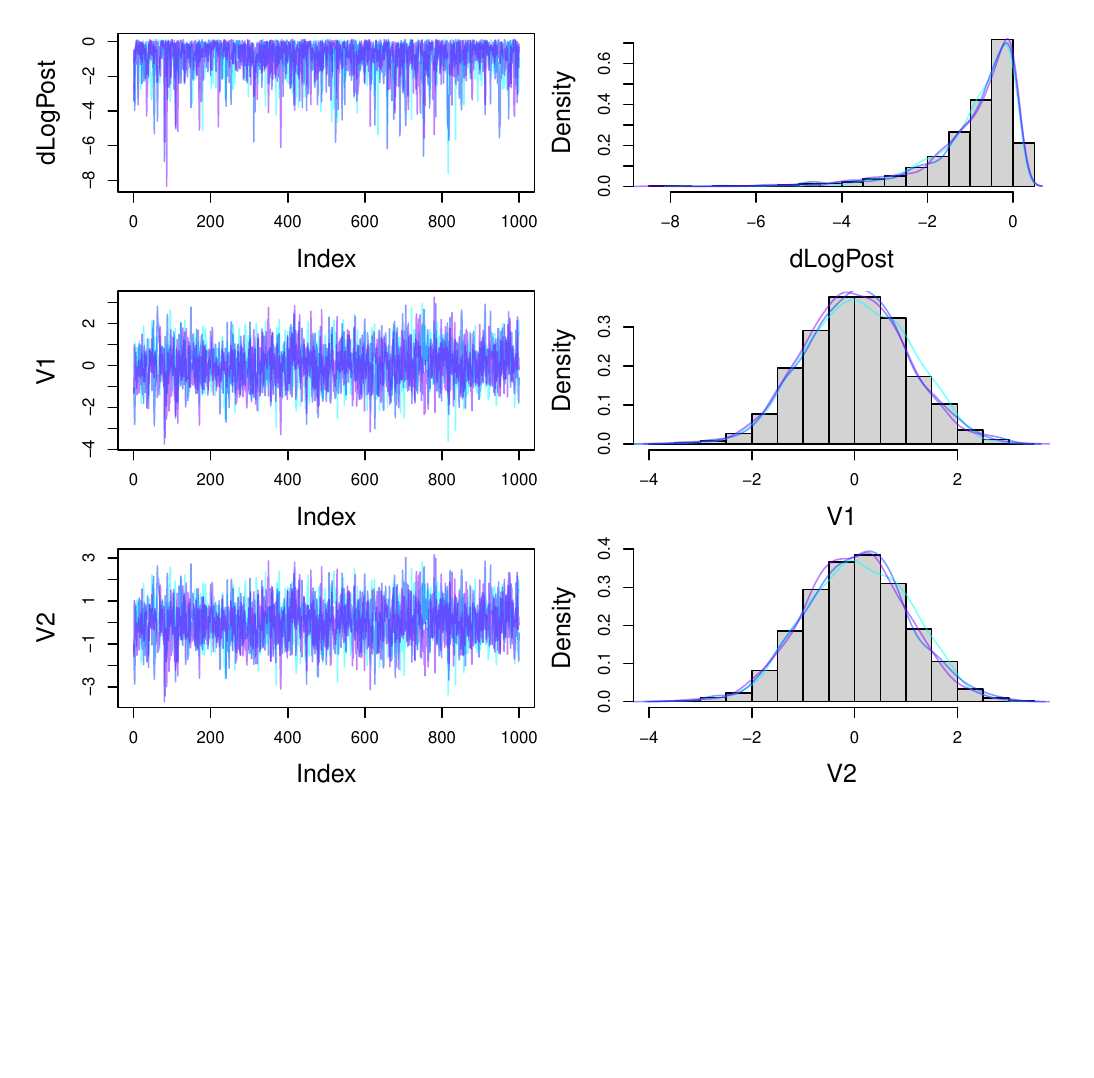}
\caption{
\textbf{Trace of downsampled AMC chains sampling the bivariate normal distribution.} 
Downsampling is performed by burning the first 2000 iterations (out of 10000) and then retaining 1000 evenly spaced samples of the burnt chain.
}
\end{center}
\end{figure}

\newpage
\section{Traces of DEMC chains fitted to the multivariate Gaussian}

\begin{figure}[H]
\begin{center}
\includegraphics[page=9, width=1\linewidth]{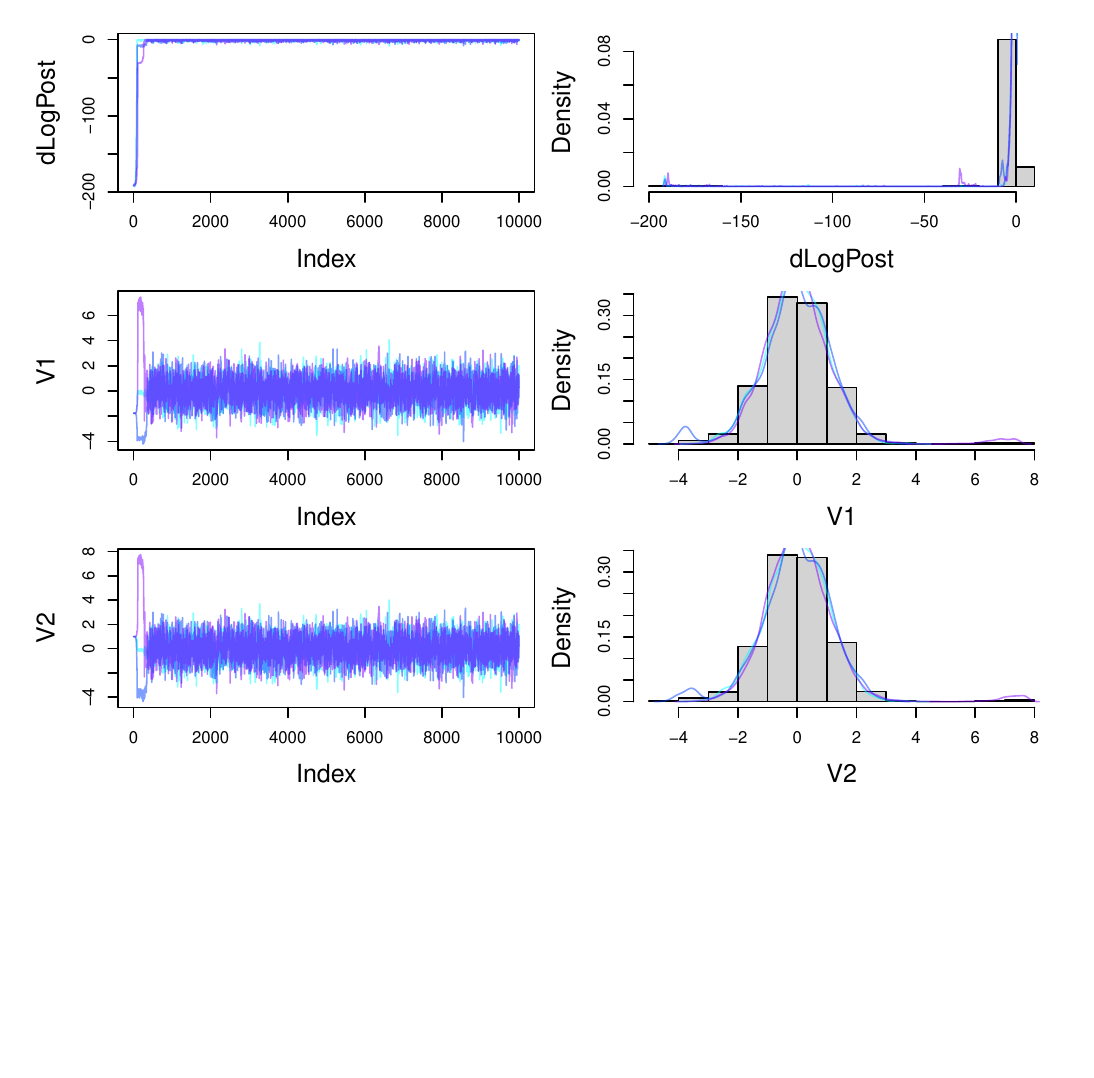}
\caption{\textbf{Trace of full DEMC chains sampling the bivariate normal model.}} 
\end{center}
\end{figure}

\begin{figure}[H]
\begin{center}
\includegraphics[page=10, width=1\linewidth]{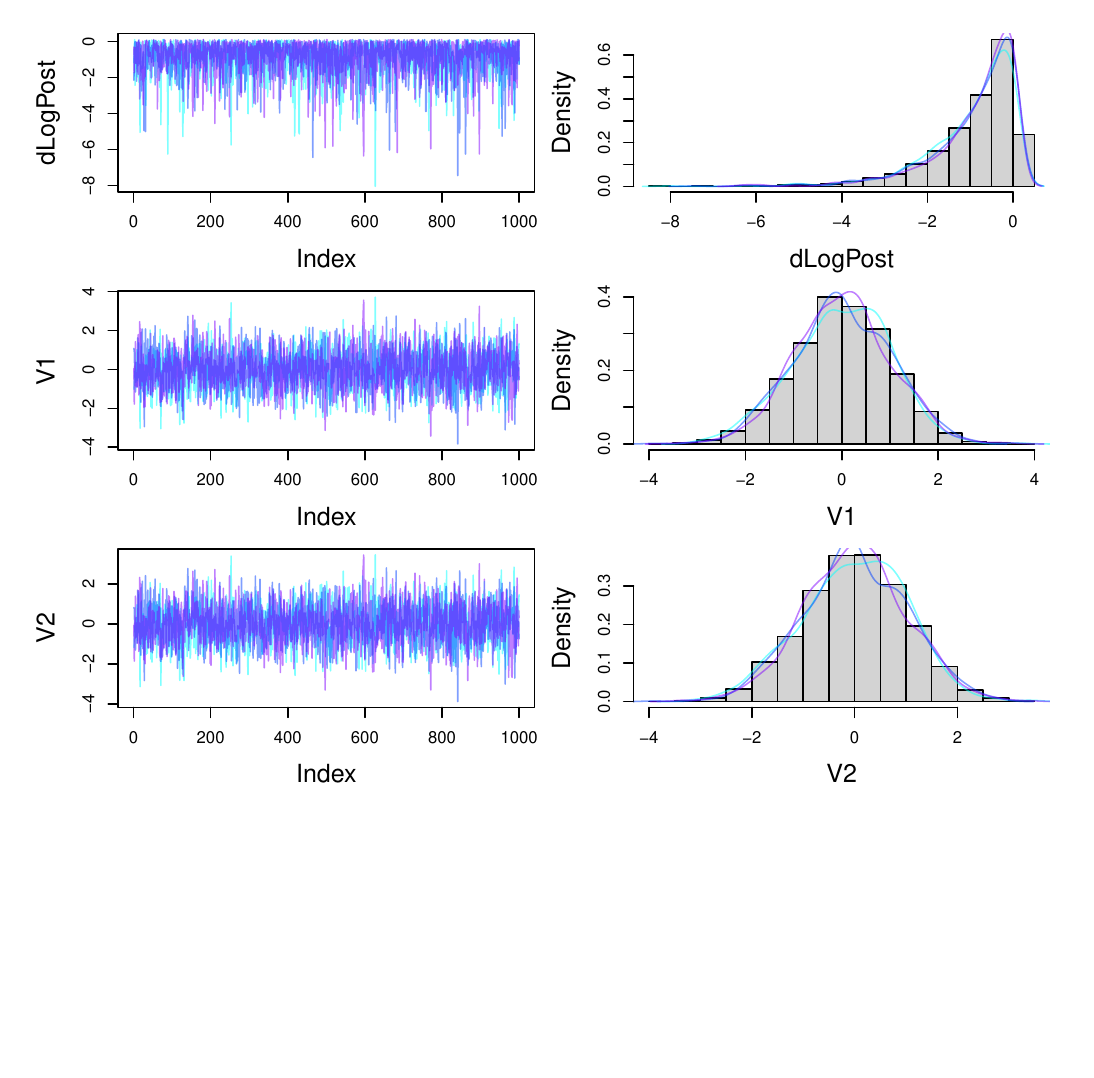}
\caption{
\textbf{Trace of downsampled DEMC chains sampling the bivariate normal distribution.} 
Downsampling is performed by burning the first 2000 iterations (out of 10000) and then retaining 1000 evenly spaced samples of the burnt chain.
}
\end{center}
\end{figure}

\newpage
\section{Traces of AMC chains fitted to the ODE system}

\begin{figure}[H]
\begin{center}
\includegraphics[page=11, width=1\linewidth]{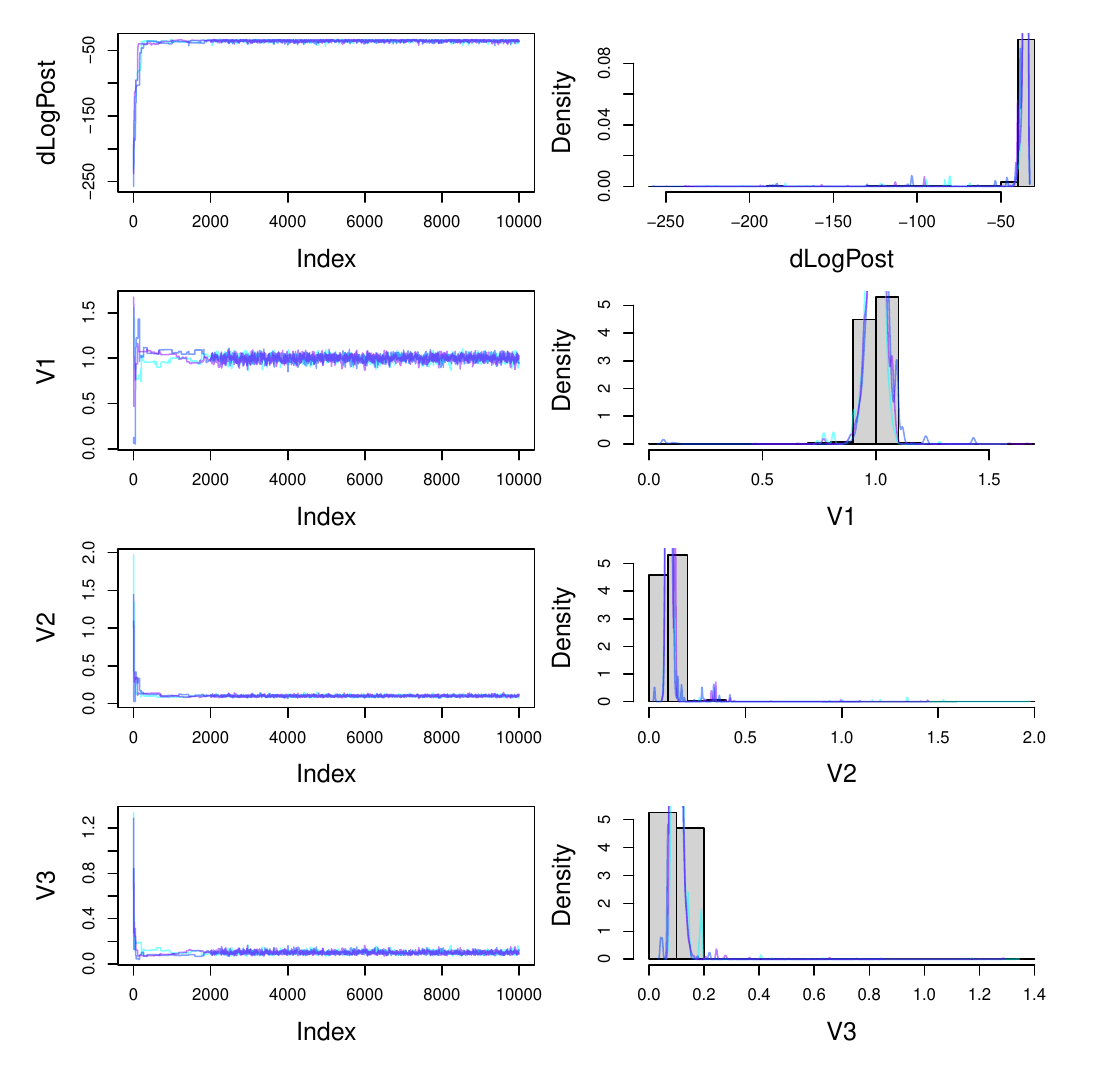}
\caption{\textbf{Trace of full AMC chains sampling the ODE model.}} 
\end{center}
\end{figure}

\begin{figure}[H]
\begin{center}
\includegraphics[page=12, width=1\linewidth]{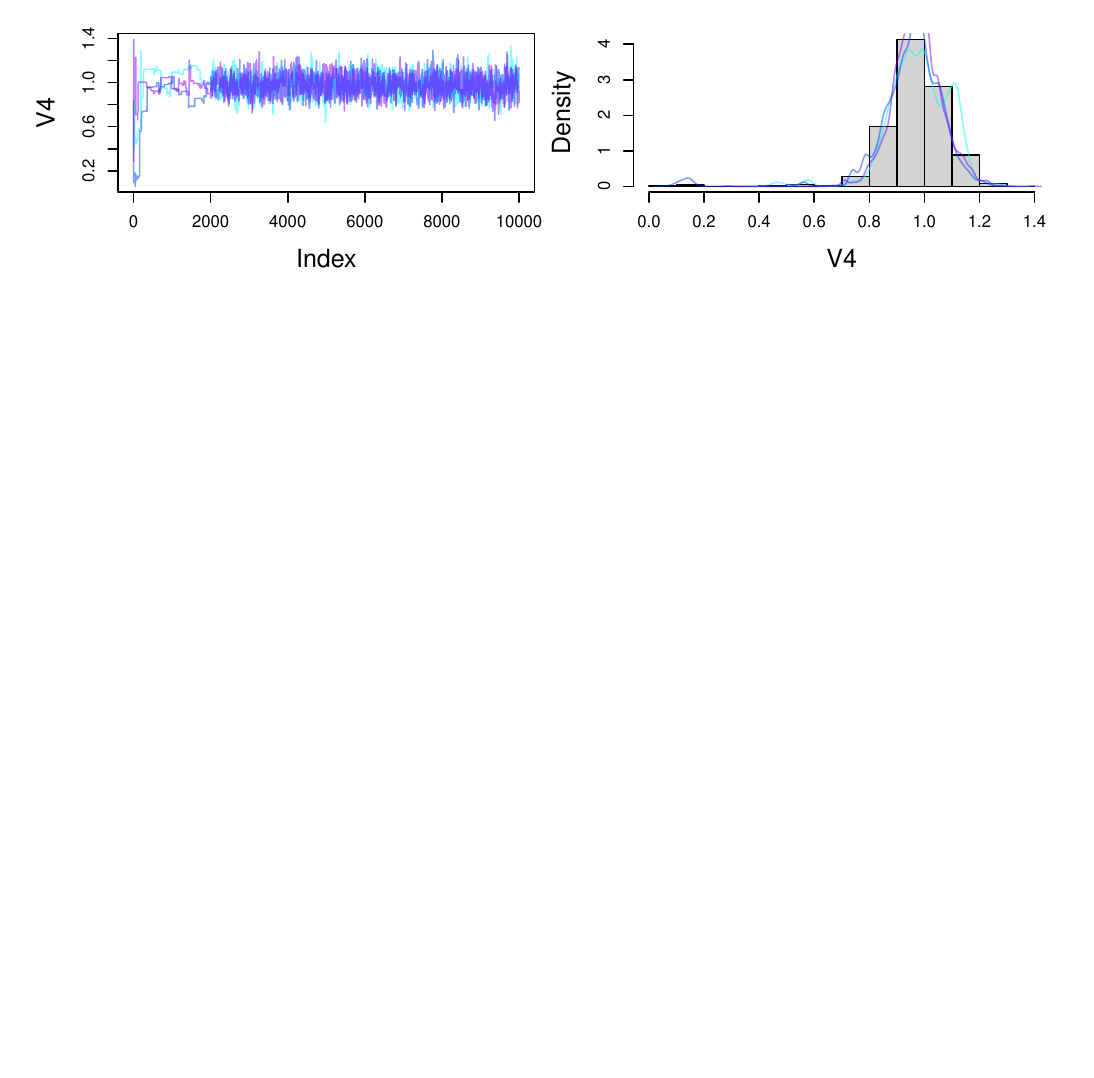}
\caption{
Continued
} 
\end{center}
\end{figure}

\begin{figure}[H]
\begin{center}
\includegraphics[page=13, width=1\linewidth]{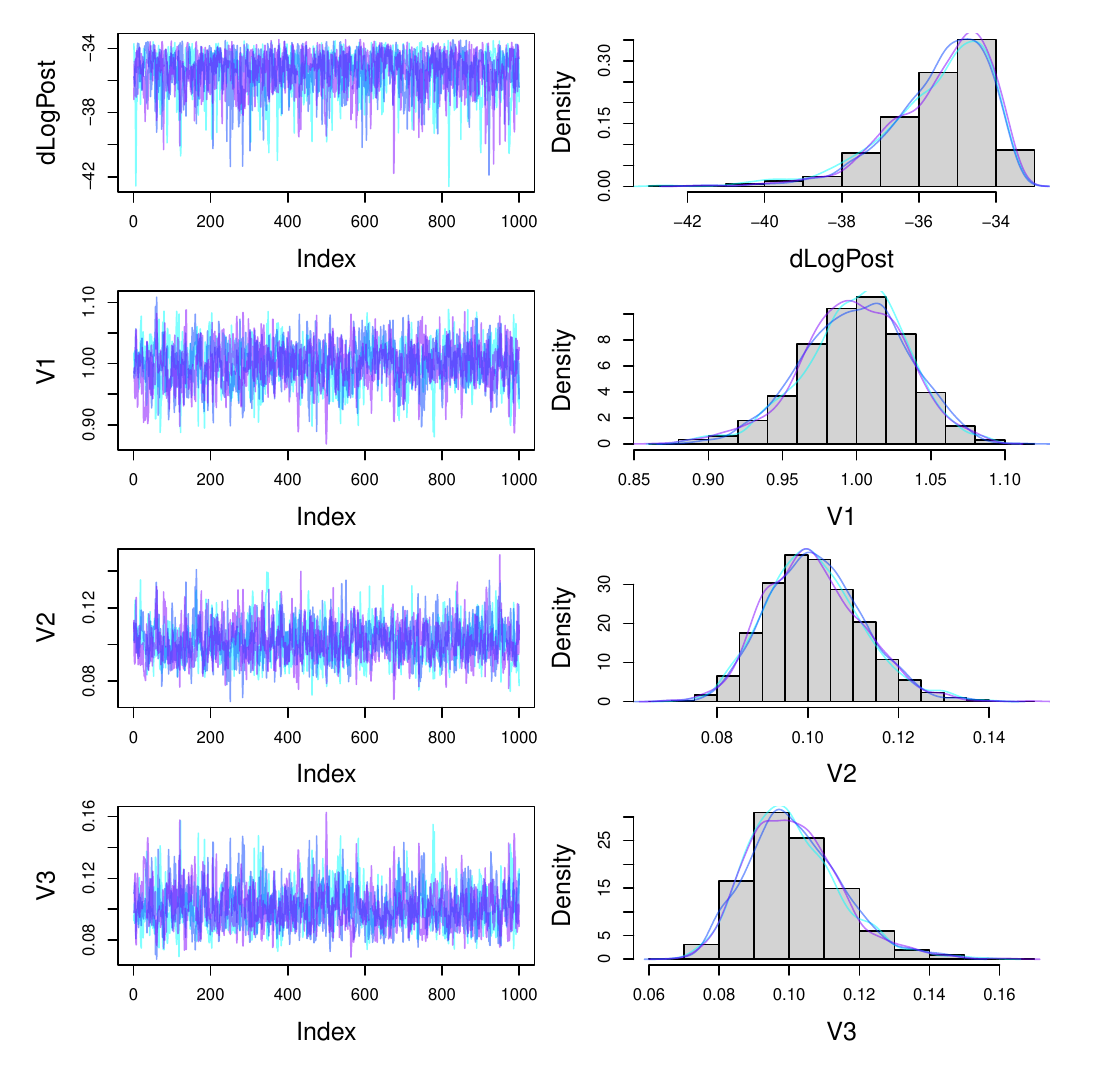}
\caption{
\textbf{Trace of downsampled AMC chains sampling the ODE model.} 
Downsampling is performed by burning the first 2000 iterations (out of 10000) and then retaining 1000 evenly spaced samples of the burnt chain.
} 
\end{center}
\end{figure}

\begin{figure}[H]
\begin{center}
\includegraphics[page=14, width=1\linewidth]{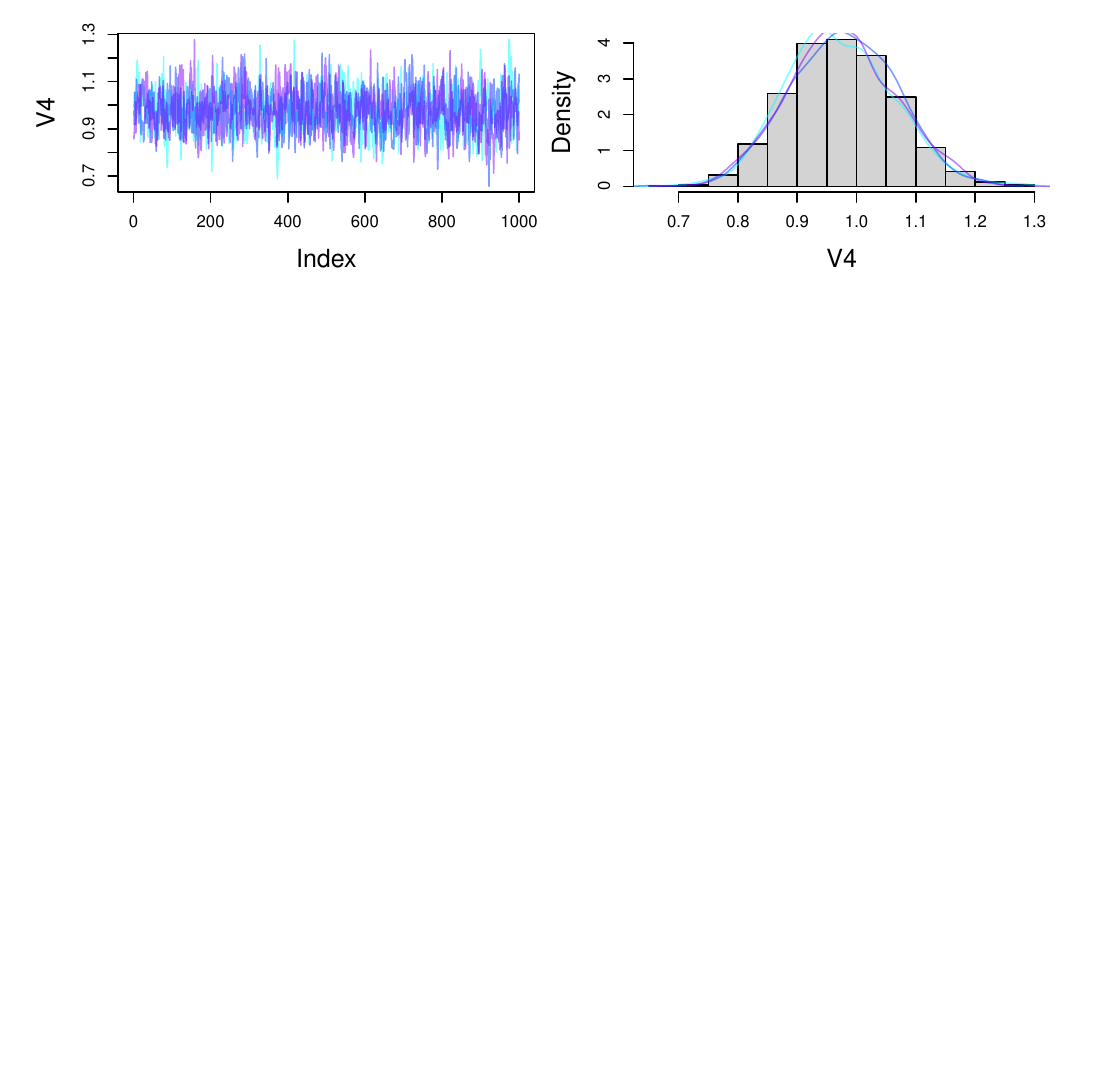}
\caption{Continued}
\end{center}
\end{figure}

\newpage
\section{Traces of DEMC chains fitted to the ODE system}

\begin{figure}[H]
\begin{center}
\includegraphics[page=15, width=1\linewidth]{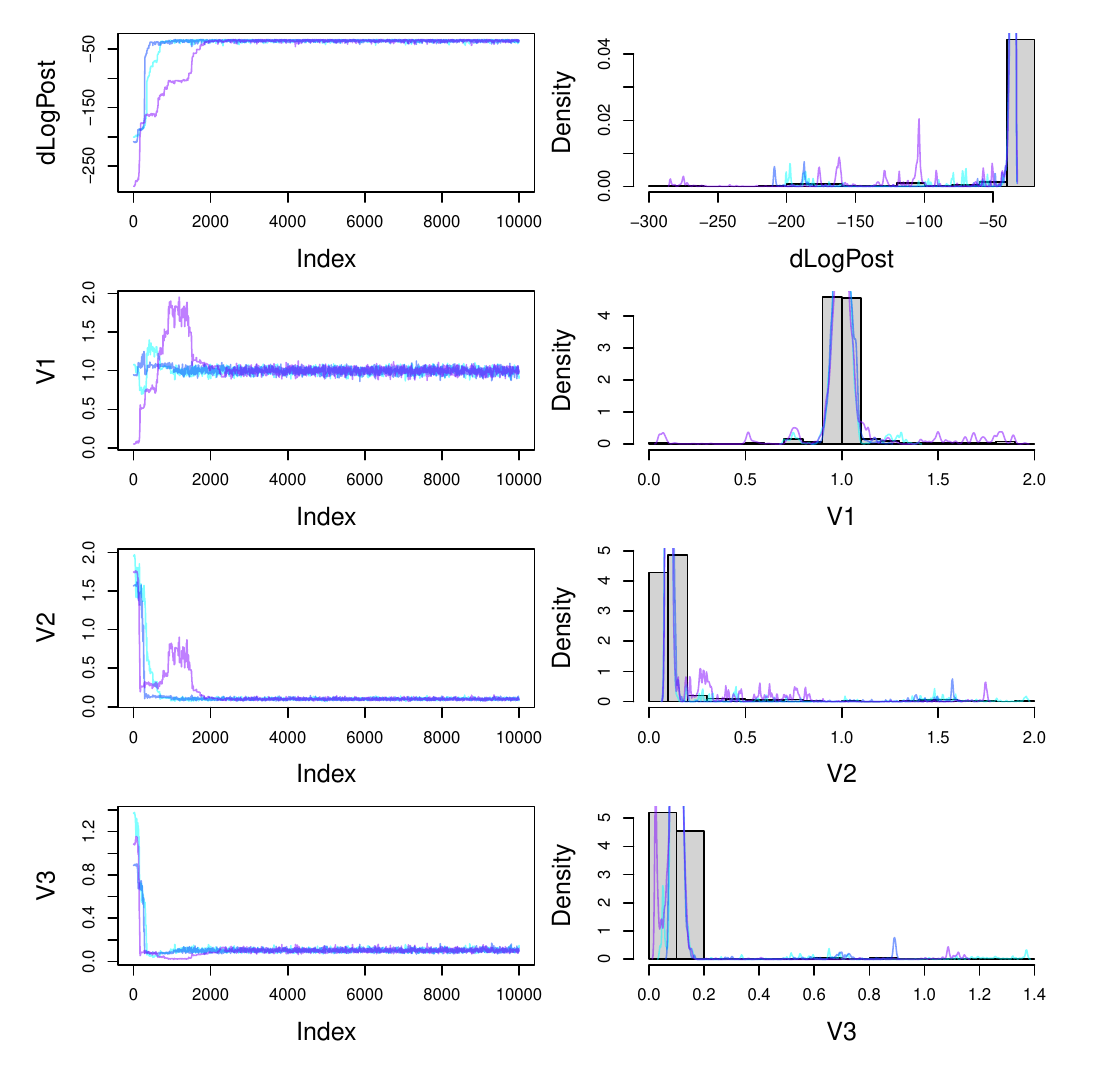}
\caption{\textbf{Trace of full DEMC chains sampling the ODE model.}} 
\end{center}
\end{figure}

\begin{figure}[H]
\begin{center}
\includegraphics[page=16, width=1\linewidth]{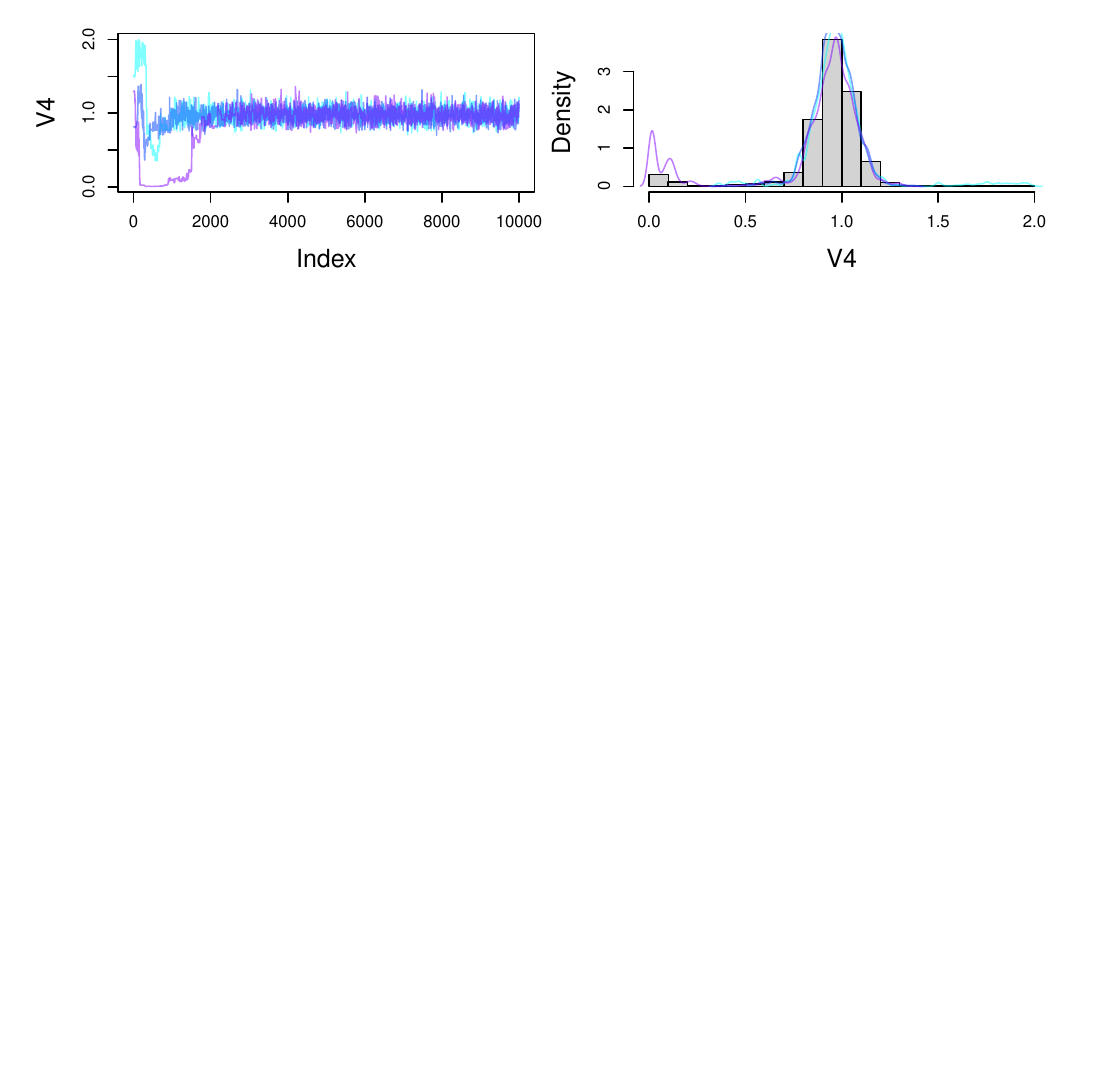}
\caption{Continued} 
\end{center}
\end{figure}

\begin{figure}[H]
\begin{center}
\includegraphics[page=17, width=1\linewidth]{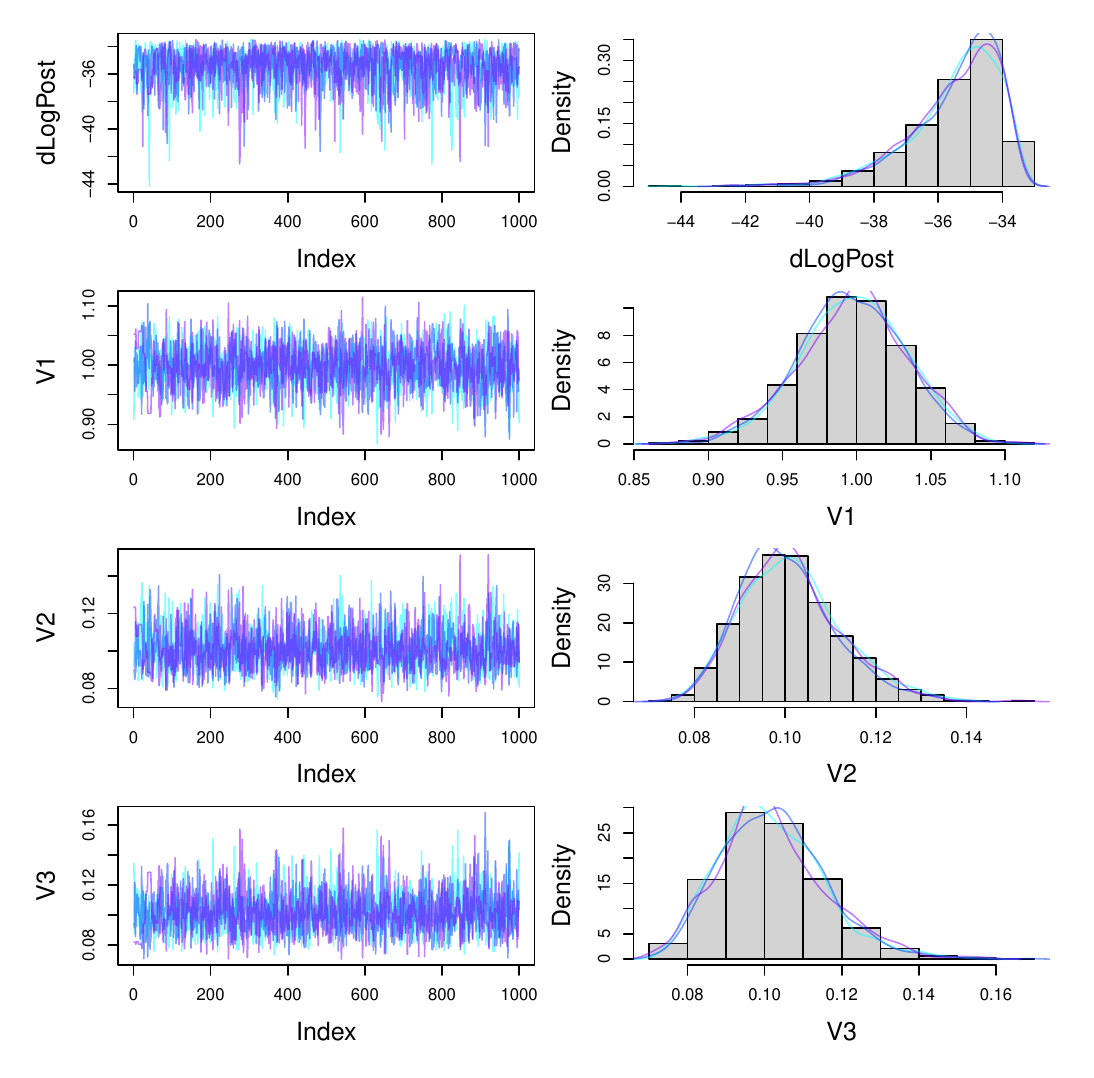}
\caption{
\textbf{Trace of downsampled DEMC chains sampling the ODE model.} 
Downsampling is performed by burning the first 2000 iterations (out of 10000) and then retaining 1000 evenly spaced samples of the burnt chain.
}
\end{center}
\end{figure}

\begin{figure}[H]
\begin{center}
\includegraphics[page=18, width=1\linewidth]{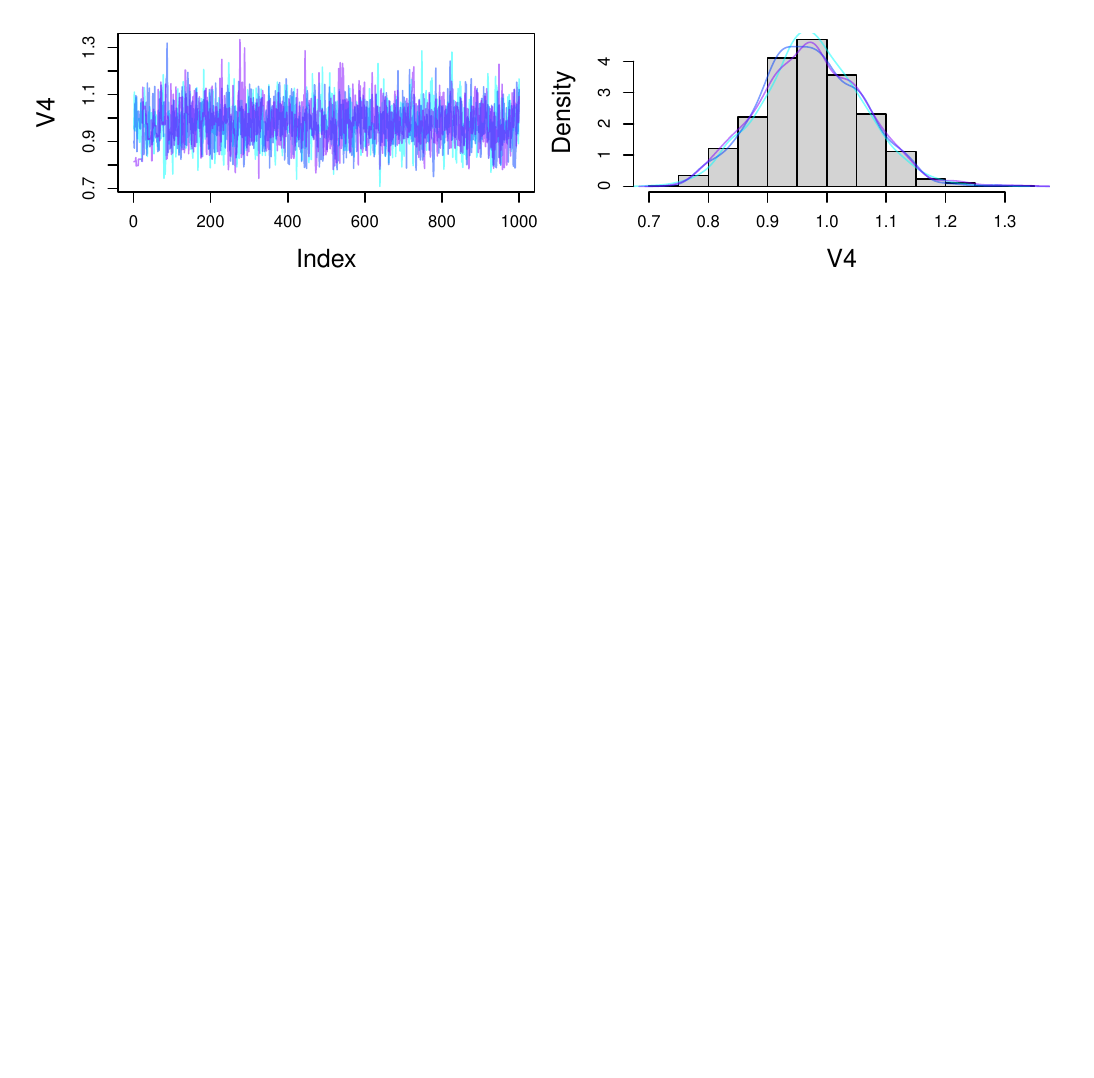}
\caption{
Continued
}
\end{center}
\end{figure}

%

\end{document}